\documentclass[12pt]{article}

\usepackage{bm}
\usepackage{float}
\usepackage{relsize}
\usepackage{array}
\usepackage{fullpage}
\usepackage{amsfonts}
\usepackage{amsmath}
\usepackage{setspace}
\usepackage{easybmat}
\usepackage{slashed}
\usepackage{amsmath}
\usepackage[mathscr]{eucal}
\usepackage{amssymb}
\usepackage{graphicx}
\usepackage{verbatim}
\usepackage{caption}
\usepackage{subcaption}
\usepackage{hyperref}

\allowdisplaybreaks
\def\timesbox{\hbox{$\scriptscriptstyle\times$}}
\def\ant{ {{\lower 1ex  \timesbox} \atop {\raise 1.5ex  \timesbox}}}
\def\f{\frac}
\def\pa{\partial}
\def\non{\nonumber\\}
\def\a{\alpha}
\def\b{\beta}

\def\d{\delta}
\def\e{\epsilon}
\def\g{\gamma}
\def\k{\kappa}
\def\l{\lambda}

\def\m{\mu}
\def\n{\nu}

\def\t{\tau}
\def\th{\theta}
\def\D{\Delta}

\def\G{\Gamma}

\newcommand{\mathsym}[1]{{}}
\newcommand{\unicode}[1]{{}}

\newcommand{\be}{\begin{equation}}
\newcommand{\ee}{\end{equation}}
\newcommand{\beqa}{\begin{eqnarray}}
\newcommand{\eeqa}{\end{eqnarray}}
\newcommand{\bsp}{\begin{split}}
\newcommand{\esp}{\end{split}}
\newcommand{\bgth}{\begin{gather}}
\newcommand{\egth}{\end{gather}}

\newcommand{\del}{\partial}

\newcommand{\tr}{\hbox{tr}}

\begin{document}

\title{\textbf{Partition functions for $U(1)$ vectors and phases of scalar QED in AdS}
\\
\author{Astha Kakkar$^{a}$\footnote{asthakakkar8@gmail.com} and 
Swarnendu Sarkar$^{b}$\footnote{swarnen@gmail.com}\\
$^a$\small{{\em Department of Physics and Astrophysics,
University of Delhi,}} \\ 
\small{{\em Delhi 110007, India}}\\
$^b$\small{{\em Department of Physics, Vidyasagar University,}}\\
\small{{\em Midnapore 721102, India}}\\
}}

\date{}

\maketitle

\abstract{We extend the computation of one-loop partition function in AdS$_{d+1}$ using the method in \cite{Kakkar:2022hub} and \cite{Kakkar:2023gzu} for scalars and fermions to the case of $U(1)$ vectors. This method utilizes the eigenfunctions of the AdS Laplacian for vectors. For finite temperature, the partition function is obtained by generalizing the eigenfunctions so that they are invariant under the quotient group action, which defines the thermal AdS spaces. The results obtained match with those available in the literature. As an application of these results, we then analyze phases of scalar QED theories at one-loop in $d=2,3$. We do this first as functions of AdS radius at zero temperature showing that the results reduce to those in flat space in the large AdS radius limit. Thereafter the phases are studied as a function of the scalar mass and temperature. We also derive effective potentials and study phases of the scalar QED theories with $N$ scalars.}

\newpage

\tableofcontents

\baselineskip=18pt

\section{Introduction}\label{introf}

Study of quantum field theories in Anti-de Sitter (AdS) spaces has quite a long history. A partial list of references in this field includes \cite{Burgess:1984ti}-\cite{Aharony:2012jf}. In the recent years there has been revival of interest in the subject. We list some of these works which constitute a growing body in the literature. A review containing connection of boundary correlation functions in AdS and non-perturbative flat space scattering amplitudes is \cite{Kruczenski:2022lot}. There have been recent works focusing on critical points of field theories in AdS which shed light on conformal field theories in flat space with boundaries \cite{Carmi:2018qzm}-\cite{Giombi:2021cnr}. Field theories with large number of flavors have been studied in \cite{Carmi:2018qzm} \cite{Ankur:2023lum} \cite{Carmi:2019ocp} \cite{Carmi:2021dsn}. \cite{Meineri:2023mps} and \cite{Lauria:2023uca} explored RG flows in AdS.

Since AdS space acts as an infrared regulator, one expects quantum field theories to be more well behaved in infrared as compared to those in flat space. This had also been another prime motivation for studying field theories in AdS. See for example \cite{Callan:1989em}. With this aspect in mind, in \cite{Kakkar:2022hub} and \cite{Kakkar:2023gzu} we have analyzed phases of various theories containing scalars and fermions in AdS which show several novel features as compared to their flat space counterparts. 
One of the aims of this paper is to  extend the analysis to theories with vector field. In particular we study the phases of scalar QED theories in AdS$_{d+1}$ (for $d=2,3$) spaces both at zero and finite temperatures at one-loop level. For this we use the effective potential which is derived by using the standard procedure of computing the one-loop determinant or the partition function. For scalars the partition function was calculated by introducing a method in \cite{Kakkar:2022hub}. Another purpose of this paper is to  utilize the same procedure as for scalars and for fermions in \cite{Kakkar:2023gzu} to compute the partition function for vectors using the Poincar\'e coordinates. In the initial parts of the paper we derive the one-loop determinant for Euclidean AdS$_{d+1}$ by finding the eigenfunctions of the Laplacian for vectors. We next study partition functions on thermal AdS$_{d+1}$ which has the conformal boundary $S^1\times S^{d-1}$. This space is the quotient space ${\mathbb H}^{d+1}/{\mathbb Z}$. To compute the finite temperature partition function we define eigenfunctions that are invariant under the action of $\g^n \in {\mathbb Z}$ including the action of $\g^n$ on the vector field. The results obtained are shown to match with those existing in the literature which are derived using other methods \cite{Gibbons:2006ij}-\cite{Gupta:2012he}.    

The zero temperature effective potential is divergent both in the ultraviolet and in infrared. The latter is due to the divergent AdS volume. These are same as those in flat space. However unlike in flat space, in AdS the thermal contribution is not proportional to the volume of the underlying space. These divergences at zero temperature are taken care of as was done in \cite{Kakkar:2022hub} and \cite{Kakkar:2023gzu}. The ultraviolet divergences in the effective potential are regularized by analytically continuing the result in the number of dimensions. The resulting expression has poles only for odd $d$. For AdS$_3$ the analytically continued expression thus gives a finite result. The effective potential for AdS$_4$ are renormalized using the minimal subtraction scheme as well as by including counterterms which are obtained by setting appropriate renormalization conditions. One further needs to regularize the volume to proceed with the analysis, the procedure for which is reviewed in \cite{Kakkar:2022hub}. Implementing these, we derive results for scalar QED theories with a single scalar as well as for theories with $N$ scalars. We have elaborately studied the phases, the results of which are summarised below.        
  
\noindent 
At zero temperature or for Euclidean AdS we study the phases as a function of AdS radius $L$, scalar charge $e$ and scalar mass $m^2$. 

AdS$_3$: The theory with a massless scalar has three dimensionful parameters, the scalar charge $e$, the $\phi^4$-coupling $\l$ and the AdS radius $L$. For the one-loop correction to be of the same order in perturbation theory as the tree-level we assume $\l \sim e^2$. 
At large values of $L$ compared to $1/e^2$ it is shown that the leading term of the potential (in $1/L$) is that of flat space. Including terms subleading in $1/L$, we get a phase plot in the $L$-$e$ plane with three distinct regions and a line of the first order transition (Figure \ref{EAdS3}). We further analyze scalar QED theories with finite number of massive scalars for fixed value of $L$. The phase diagrams, shown in Figure \ref{EAdS3N} have similar regions as in Figure \ref{EAdS3}.  

AdS$_4$: We renormalize the potential using the $\overline{\text{MS}}$ scheme and introducing a renormalization scale $\m$. 
The analysis is first done in the massless scalar limit (in the Landau gauge) which enables us to compare with the Coleman-Weinberg potential in the flat space limit setting $\l \sim e^4$. For this massless theory, we first analyze two regimes analytically, large $L$ and small $\phi_{cl}$, that gives insight into the behaviour of the effective potential. In the $L\rightarrow \infty$ limit we show that we recover the flat space symmetry broken phase with a the double well potential. As $L$ is decreased the double well goes over to a potential with a single minimum through a first order transition at a critical radius $L_{c1}$. For the small values of $\phi_{cl}$, again by varying $L$, we have another first order transition at $L_{c2}$ where the minima exchange dominance. In this entire analysis the charge $e$ only plays the role of scaling the potential plots. A numerical analysis of the full effective potential confirms the occurrence of two first order transitions (Figure \ref{PAdS4v} (a)-(d)). For the massive theory at fixed value of $L$ we set the ultraviolet renormalization conditions at $\phi_{cl}=0$. These renormalization conditions imply that new minima exist only when the tree-level scalar mass is negative. This conclusion also holds for massive theories with $N$ scalars.

We next study the phases as a function of scalar mass-squared, $m^2$ and the inverse temperature, $\b$ setting $L=1$. 

AdS$_3$: The phase plot is divided into two regions separated by a vertical line at $\b=\b_1$. This line represents the second order phase transition in the theory. There is further a first order transition beyond this value of $\b$ (Figure \ref{TAdS3}). Due to the negative sign of the regularized volume for AdS$_3$ the $U(1)$ symmetry remains broken at high temperatures as in the case of theories with only scalars \cite{Kakkar:2022hub}. Increasing the number of scalars give qualitatively the same phase plot with the phase-regions being scaled. 

AdS$_4$: The renormalization condition defining the finite coupling $\l$ is set at $\phi_{cl}$= 0. In this case for zero temperature, $L$ is the only scale in the problem and shape of the potential remains invariant with change in $L$.  The large $\phi_{cl}$ limit then gives the flat space result where leading order log term matches with the Coleman-Weinberg potential \cite{Coleman:1973jx}\cite{Weinberg:1973am}. The difference in the subleading terms in $\phi_{cl}$ arises since we renormalize the effective potential in AdS at $\phi_{cl}=0$.  For finite temperature with $L = 1$ the phase transitions occur only when $m^2$, which is the other dimensionful parameter, is nonzero and negative. We get a $\b - m^2$ phase plot with four different regions where the region boundaries are characterized by a first order transition, a second order transition and separation of a region with only a single minima at $\phi_{cl}=0$ (Figure \ref{TAdS4}). For finite $N$ the phase plots are similar to those discussed above.

The paper is organized as follows. After presenting the basic setup in the context of scalar QED in section \ref{partv} we derive the contribution to the partition function from  $U(1)$ vectors in section \ref{comptrv}. In section \ref{zerotvtr} we compute the one loop vector determinant on Eucliedean AdS$_{d+1}$. The computation is then extended to that of thermal AdS using the method introduced in \cite{Kakkar:2022hub}, in section \ref{thermal3v} for AdS$_3$ and in section \ref{thermaldv} for general $d$ dimensions. Some details of these computations for zero and finite temperature partition functions are given in appendices \ref{detailsv} and \ref{acomptv}. In section \ref{phasesssqed} we study the phases of single-scalar QED. The theories in AdS$_3$ and in AdS$_4$ are studied in sections \ref{phases3v} and in section \ref{phases4v} respectively. We discuss the gauge independence of the vacuum energy density in section \ref{gdep}. Effective potential for QED theory with $N$ scalars is derived in section \ref{effpotNs} and the phases are analyzed. The detailed effective potentials for single scalar and $N$-scalars in AdS$_4$ are written down in appendix \ref{effads4v}. We conclude this paper with a discussion on some possible future directions in section \ref{summaryv}.

\section{Basic setup for scalar QED}\label{partv}

In this section, we highlight the general setup for scalar quantum electrodynamics. We begin with the following Euclidean Lagrangian

\beqa
\mathcal{L}_E=\frac{1}{4}F^{}_{\mu \nu}F^{\mu \nu}+|D_{\mu}\phi|^{2}+V(\phi)
\eeqa
where the potential is
\beqa\label{clpot}
V(\phi)=m^{2} \phi^* \phi + \frac{\lambda}{4}  (\phi^* \phi)^2~~~~~\mbox{and}~~~~~
D_{\mu}=\del_\mu+i e A_\mu.
\eeqa
To proceed with the analysis we write the scalar field as

\beqa
\phi=\frac{1}{\sqrt{2}}\left(\phi_1 +i\phi_2\right)
\eeqa
and then expand about the classical vacuum expectation value $\phi_{cl}$ as
\begin{eqnarray}
\phi_1=\sqrt{2}\phi_{cl} + \eta_1 
~~~~~\mbox{and}~~~~~ \phi_2=\eta_2.
\end{eqnarray}

This gives us the following Lagrangian quadratic in fields
\begin{eqnarray}
\mathcal{L}_E&=&\frac{1}{4}F^{}_{\mu \nu}F^{\mu \nu}+\frac{1}{2}(\del_{\mu}\eta_1)^2+\frac{1}{2}(\del_{\mu}\eta_2)^2+\frac{1}{2}\eta^{2}_{1}\left(m^2+\frac{3 \l}{2}\phi^{2}_{cl}\right)+\frac{1}{2}\eta^{2}_{2}\left(m^2+\frac{\l}{2}\phi^{2}_{cl}\right) \non
&+&m^2\phi_{cl}^2+\frac{\l}{4}\phi^{4}_{cl}+ e^2 A_{\mu}A^{\mu}\phi^{2}_{cl}+\sqrt{2}(\del_\mu \eta_2)e A^\mu \phi_{cl}.\label{mixedterm}
\end{eqnarray}

Including the source terms $J_1\phi_1+J_2\phi_2$ in ${\cal L}$, the linear term in $\eta_1$ vanish due to $\d{\cal L}/\d \phi_{1}+J_1=0$, evaluated at $\phi_1=\sqrt{2}\phi_{cl}$ and $\phi_2=0$. In other words, the background values of $\phi_1$ and $\phi_2$ satisfy the classical equation of motion. In the following however, we shall be working with $J_1=J_2=0$. The interaction terms are omitted as in this paper we shall be restricting to a one-loop analysis.  The infinitesimal local symmetry transformation resulting from the gauge symmetry, $\phi \rightarrow e^{-i \alpha(x)} \phi$, in terms of the shifted fields is
\begin{eqnarray}
\delta\eta_1= -\alpha(x)\eta_2 ~~~;~~~\delta\eta_2= \alpha(x)(\phi_{cl}+\eta_1) ~~~;~~~\delta A_\mu=-\frac{1}{e}\del_\mu \alpha(x).
\end{eqnarray}

Using the standard method (see for example \cite{Peskin:1995ev}) of including a gauge fixing term $G^2/2$ in the $R_{\xi}$ gauge which removes the mixed term in (\ref{mixedterm}) with 

\begin{equation}
G=\frac{1}{\sqrt{\xi}}(\nabla_{\mu}A^{\mu}+\sqrt{2}\xi e \eta_2 \phi_{cl})
\end{equation}

and including the corresponding Fadeev-Popov Ghosts we have
\begin{eqnarray}\label{scalarqedL}
\mathcal{L}_E&=&{\cal L}_{\mbox{\tiny{gauge}}}+{\cal L}_{\mbox{\tiny{ghost}}}+{\cal L}_{\mbox{\tiny{scalars}}}\\
&=&\frac{1}{4}F^{}_{\mu \nu}F^{\mu \nu}+\f{1}{2\xi}(\nabla_{\m}A^{\m})^2+\f{1}{2}M_v^2 A_{\mu}A^{\mu}+\bar{c}\left(-\nabla^2+2 \xi e^2 \phi^{2}_{cl}\right)c \label{scalarqedL1}\\ \nonumber
&+&\frac{1}{2}(\del_{\mu}\eta_1)^2+\frac{1}{2}(\del_{\mu}\eta_2)^2+\frac{1}{2}M^{2}_{1}\eta^{2}_{1}+\frac{1}{2}M^{2}_{2}\eta^{2}_{2}+m^2\phi_{cl}^2+\frac{\l}{4}\phi^{4}_{cl}
\end{eqnarray}

where

\begin{eqnarray}\label{massdefv}
M^{2}_{1}=m^2+\frac{3 \l}{2}\phi^{2}_{cl}~~~~M^{2}_{2}=m^2+\frac{\l}{2}\phi^{2}_{cl}+2 \xi e^2 \phi^{2}_{cl} ~~~~~
M^{2}_{v}=2e^2 \phi^{2}_{cl}.
\end{eqnarray}

The one loop partition function in the absence of sources is 

\beqa
Z^{(1)}=\int{\cal D}A_{\m}{\cal D}c{\cal D}\bar{c}{\cal D}\eta_1{\cal D}\eta_2e^{-S_E}=Z^{(1)}_{\mbox{\tiny{gauge}}}Z^{(1)}_{\mbox{\tiny{ghost}}}Z^{(1)}_{\mbox{\tiny{scalars}}}.
\eeqa

For the computation of $Z_{\mbox{\tiny{gauge}}}$, it is convenient to decompose the gauge field into transverse and longitudinal modes $A_{\m}=A_{\m}^T+A_{\m}^L$ so that $\nabla^{\m}A_{\m}^T=0$ and $A_{\m}^L=\pa_{\mu}\chi$. The $F_{\m\n}^2$ term is independent of $A_{\mu}^L$. The longitudinal mode gets contribution from the gauge-fixing $(\nabla_{\m}A^{\m})^2/(2\xi)$ and the vector mass terms.
The transverse and the longitudinal contributions to the action thus decouple and the gauge partition function is then

\beqa
Z^{(1)}_{\mbox{\tiny{gauge}}}=\int{\cal D}A_{\m}^T{\cal D}A_{\m}^Le^{-S{\mbox{\tiny{gauge}}}}=Z_{\mbox{\tiny{gauge}}}^TZ_{\mbox{\tiny{gauge}}}^L.
\eeqa

The effective potential thus is

\beqa
\label{potsqed}
V^{}_{eff}(\phi^{}_{cl})&=&-\frac{1}{{\cal V}^{}_{d+1}}\left[\log Z_{\mbox{\tiny{gauge}}}^{(1)}+\log Z_{\mbox{\tiny{ghost}}}^{(1)}+\log Z_{\mbox{\tiny{scalars}}}^{(1)}\right]+V(\phi_{cl})\non
&=&\f{1}{{\cal V}_{d+1}}\left[\f{1}{2}\mbox{tr}\log[-\mathcal{O}^T+M_v^2-d+1]+\f{1}{2}\mbox{tr}\log[-\mathcal{O}^L+\xi M_v^2]-\mbox{tr}\log[-\square^{}_{E}+\xi M_v^2]\right.\non
&+&\left.\frac{1}{2}\mbox{tr} \log[-\square^{}_{E}+M_1^2]+\frac{1}{2}\mbox{tr} \log[-\square^{}_{E}+M_2^2]\right]+V(\phi^{}_{cl}).
\eeqa

${\cal V}_{d+1}$ is the volume of the $(d+1)$-dimensional space. $\square_E$ is the scalar Laplacian. ${\cal O}^T$ and ${\cal O}^L$ are the transverse and longitudinal operators acting on vectors respectively. The longitudinal contribution is same as that of a scalar of mass $\sqrt{\xi}M_v$. This will be verified in section \ref{zerotvtr}. The ghost contribution, which is also like scalar field with the same mass has the effect of canceling the longitudinal contribution of the vector field and we are left with just the transverse contribution minus a massive scalar contribution having mass $\sqrt{\xi}M_v$.

The $\log$ of the traces can be obtained (up to a constant independent of $\phi_{cl}$) by integrating the following with respect to the mass-squared: 

\beqa\label{trv}
\frac{1}{2{\cal V}^{}_{d+1}}\mbox{tr}\left[ \frac{1}{-\square^{}_{E}+M^{2}_{1,2}(\phi^{}_{cl})}\right]~~~~~~~\mbox{and}~~~~~~~\frac{1}{{2 \cal V}^{}_{d+1}}\mbox{tr}\left[ \frac{1}{-\mathcal{O}^T+M_v^2(\phi_{cl})-d+1}\right].
\eeqa

For theories with only scalars in AdS$_{d+1}$, the scalar trace was computed and phases of various theories involving only scalars were studied in \cite{Kakkar:2022hub}. Like the scalar the result for vector trace is known in the literature as mentioned in the introduction. In the following few sections, working in Poincar\'e coordinates, we reproduce these results for the trace (\ref{trv}) for the vectors at zero and finite temperature using an alternate method similar to that of scalars in \cite{Kakkar:2022hub} and spinors in \cite{Kakkar:2023gzu}.

\section{Computation of traces for vectors}\label{comptrv}

In the following sections we compute the trace (\ref{trv}) for vectors and the corresponding partition functions on Euclidean and thermal AdS spaces. The method uses eigenfunctions of the Laplace operators for vectors on Euclidean AdS, ${\mathbb H}^3$ using Poincar\'e coordinates. These functions are then generalized for thermal AdS (${\mathbb H}^3/{\mathbb Z}$), so that they are invariant under thermal identifications. The results obtained are then used to study the phases of scalar QED theories in later sections. 

\subsection{Zero temperature: Euclidean AdS}\label{zerotvtr}

The $U(1)$ gauge field part of the Lagrangian (\ref{scalarqedL}) is

\begin{eqnarray}\label{gaugeL}
\mathcal{L}_{\mbox{\tiny{gauge}}}=\frac{1}{4}F^{}_{\mu \nu}F^{\mu \nu}+\f{1}{2\xi}(\nabla_{\m}A^{\m})^2+\f{1}{2}M_v^2 A_{\mu}A^{\mu}.
\end{eqnarray}

For the Euclidean AdS metric 

\beqa
ds^2=\f{L^2}{y^2}\left(dy^2+\eta_{\mu\nu}dx^{\mu}dx^{\nu}\right)
\eeqa

the vector field action in the Feynman gauge ($\xi=1$) can be written in the $2\times 2$ form as 

\beqa
S_g=\frac{1}{2}\int d^{d+1}x\sqrt{g}\left[\left(\begin{array}{cc}
\tilde{A}_y&\tilde{A_i}
\end{array}\right)
\left(\begin{array}{cc}
-{\cal O}+M_v^2&-2y\pa_i\\
2y\pa_i&-{\cal O}+M_v^2+(1-d)
\end{array}\right)
\left(\begin{array}{c}\tilde{A}_y\\ \tilde{A}_i\end{array}\right)\right]
\eeqa

\beqa\label{sgdef}
{\cal O}=y^2(\pa_y^2+\pa_i^2)+y(1-d)\pa_y ~~~~,~~~~M_v^2=2e^2\phi_{cl}^2~~~~~\mbox{and}~~~~~\tilde{A}_{\m}=yA_{\m}.
\eeqa

For the transverse modes which satisfy $\nabla^{\m}A_{\m}^T=0$, the action is

\beqa\label{sgT}
S_{gT}&=&\frac{1}{2}\int d^{d+1}x\sqrt{g}\left[\left(\begin{array}{cc}
\tilde{A}_y^T&\tilde{A_i^T}
\end{array}\right)
\left(\begin{array}{cc}
-{\cal O}^{\prime}+M_v^2+(1-d)&0\\
2y\pa_i&-{\cal O}+M_v^2+(1-d)
\end{array}\right)
\left(\begin{array}{c}\tilde{A}_y^T\\ \tilde{A}_i^T\end{array}\right)\right]\non
&=&\frac{1}{2}\int d^{d+1}x\sqrt{g} \left[-\tilde{A}_{\m}^T{\cal O}_{\m\n}^T\tilde{A}_{\n}^T+(M_v^2+1-d)\tilde{A}_{\m}^T\tilde{A}_{\m}^T\right]
\eeqa

where $M_v^2$ and ${\cal O}$ are defined in (\ref{sgdef}) and

\beqa
{\cal O}^{\prime}&=&y^2(\pa_y^2+\pa_i^2)-y(1+d)\pa_y+(1+d) .
\eeqa

In order to compute the one-loop determinant we need find the eigenfunctions of the transverse vector matrix operator in (\ref{sgT}). The corresponding eigenvalue equation we solve is

\beqa\label{eigent}
\left(\begin{array}{cc}
-{\cal O}^{\prime}&0\\
2y\pa_i&-{\cal O}
\end{array}\right)
\left(\begin{array}{c}\tilde{A}_y^T\\ \tilde{A}_i^T\end{array}\right)
=\left[\l^2+\left(\f{d}{2}\right)^2\right]
\left(\begin{array}{c}\tilde{A}_y^T\\ \tilde{A}_i^T\end{array}\right)
\eeqa

Like scalars the eigenvalues $\l$ are real, continuous and lie in the range $0\le\l\le\infty$. The equation (\ref{eigent}) leads to the following set of coupled differential equations     

\beqa\label{diffeigen}
-{\cal O}^{\prime}\tilde{A}^T_y=[\l^2+(d/2)^2]\tilde{A}^T_y ~~~~\mbox{and}~~~~~ -{\cal O}\tilde{A}^T_i+2i\pa_i\tilde{A}^T_y=[\l^2+(d/2)^2]\tilde{A}^T_i.
\eeqa

Solution to a similar set of equations was worked out in \cite{Mueck:1998iz}. Writing $\phi_{k,\l}(\vec{x},y)=(ky)^{d/2}K_{i\l}(ky)e^{i\vec{k}.\vec{x}}$ where $K_{i\l}(ky)$ is the modified Bessel of second kind, a solution to these coupled equations are given by\footnote{$A_y^T$ satisfies the modified Bessel equation with solution $(ky)^{d/2}K_{i\l}(ky)e^{i\vec{k}.\vec{x}}$ . Both Bessel functions of the first and second kind $I_{i\l}$ and $K_{i\l}$ are solutions of the equation. The latter is chosen as it is regular in the interior $(y \rightarrow \infty)$ of AdS. The solution to the $A_i^T$ equation is obtained by noting the following
\[
\left({\cal O}+[\l^2+(d/2)^2]\right)(ky)^{d/2+1}K_{i\l+1}(ky)e^{i\vec{k}.\vec{x}}=-2(ky)^{d/2+2}K_{i\l}(ky)e^{i\vec{k}.\vec{x}}
.\]
},

\beqa\label{trans1}
\left(\begin{array}{c}\tilde{A}_y^T\\ \tilde{A}_i^T\end{array}\right)_{(1)}&=&\f{1}{\mathscr{N}}\left(\begin{array}{c}(ky)\\ \f{ik_i}{k}\left[y\pa_y+(1-d)\right]\end{array}\right)\phi_{k,\l}(\vec{x},y)\non &=&\f{1}{|\a_T|}\left(\begin{array}{c}(ky)K_{i\l}(ky)\\ \f{ik_i}{k}\left[\a_T K_{i\l}(ky)-(ky)K_{i\l+1}(ky)\right]\end{array}\right)(ky)^{d/2}e^{i\vec{k}.\vec{x}}
\eeqa

where $\a_T=(i\l-d/2+1)$ and $\mathscr{N}$ in the first line is the normalization. It can be verified that (\ref{trans1}) satisfies $\nabla^{\m}A_{\m}^T=y\pa_{\m}\tilde{A}_{\m}^T-d\tilde{A}_y^T=0$.  

The other $(d-1)$ orthogonal, transverse solutions are obtained by setting $A_{y}^T=0$ in (\ref{diffeigen}). The $\tilde{A}_i^T$ then satisfy the modified Bessel equation, with solution $\phi_{k,\l}(\vec{x},y)=(ky)^{d/2}K_{i\l}(ky)e^{i\vec{k}.\vec{x}}$. The vectors must satisfy the transverse condition $\pa_i A^T_i=0$. With all these, for example for AdS$_3$ we have

\beqa\label{trans23}
\left(\begin{array}{c}\tilde{A}_y^T\\ \tilde{A}_1^T\\\tilde{A}_2^T\end{array}\right)_{(2)}=\f{1}{\mathscr{N}}\left(\begin{array}{c}0\\k_2\\ -k_1\end{array}\right)\phi_{k,\l}(\vec{x},y)
\eeqa

where the normalization $\mathscr{N}$ for the above vector is $k$. For AdS$_4$, in addition to (\ref{trans1}), an orthogonal set of vectors is,

\beqa\label{trans24}
\left(\begin{array}{c}\tilde{A}_y^T\\ \tilde{A}_1^T\\\tilde{A}_2^T\\ \tilde{A}_3^T\end{array}\right)_{(2)}=\f{1}{\mathscr{N}}\left(\begin{array}{c}0\\k_2\\ -k_1\\0\end{array}\right)\phi_{k,\l}(\vec{x},y)~~~~~,~~~~~
\left(\begin{array}{c}\tilde{A}_y^T\\ \tilde{A}_1^T\\\tilde{A}_2^T\\ \tilde{A}_3^T\end{array}\right)_{(3)}=\f{1}{\mathscr{N}}\left(\begin{array}{c}0\\k_1\\ k_2\\-(k_1^2+k_2^2)/k_3\end{array}\right)\phi_{k,\l}(\vec{x},y)
\eeqa

which are also orthogonal to (\ref{trans1}). In the similar manner one can write down the additional orthogonal $(d-1)$ transverse vectors for AdS$_{d+1}$.



Writing $[\psi^{T}_{\vec{k},\l}(\vec{x},y)]_{(I)}^t= \left(\begin{array}{cc}
\tilde{A}^T_y(\vec{x}, y)&\tilde{A}^T_i(\vec{x}, y)
\end{array}\right)_{(I)}$, where $I=1,\cdots, d$ and $t$ stands for the transpose, the normalization for each of the transverse vectors are worked out as (see appendix (\ref{normav}) for details)

\beqa\label{normt}
\int d^{d+1}x\sqrt{g}~[\psi^{T}_{\vec{k}^{\prime},\l^{\prime}}(\vec{x}, y)]^{\dagger}_{(I)}[\psi^T_{\vec{k},\l}(\vec{x}, y)]_{(I)}=
(2\pi)^d k^d\d^d(\vec{k}^{\prime}-\vec{k}) \d(\l^{\prime}-\l)/\m(\l)
\eeqa

where there is no summation over $I$ on the l.h.s. of (\ref{normt}) and

\beqa\label{mudef}
\m(\l)=\f{2}{\pi|\G(i\l)|^2}=2\l\sinh(\pi\l)/\pi^2.
\eeqa

The transverse trace (\ref{trv}) is now written as

\beqa
\tr\left[\frac{1}{-{\cal O}^T+M_v^2-d+1}\right]&=&\sum_{I=1}^d\int d^{d+1}x\sqrt{g}\int \f{d^dk}{(2\pi)^d}\f{1}{k^d}\int_{0}^{\infty}\f{d\l ~\m(\l)}{\l^2+\n^2_v}[\psi^{T}_{\vec{k},\l}(\vec{x},y)]^{\dagger}_{(I)}[\psi^T_{\vec{k},\l}(\vec{x},y)]_{(I)}\non\label{trzerotv1t}\\
&=& \frac{{\cal V}_{d+1}}{(4\pi)^{(d+1)/2}\G\left(\frac{d+1}{2}\right)}\int_{-\infty}^{\infty}\f{d\l }{\l^2+\n^2_v}\frac{|\G\left(\frac{d-2}{2}+i\l\right)|^2}{|\G\left(i\l\right)|^2}\non
&\times& \left[\underbrace{\l^2+\left(d/2\right)^2+(d-1)^2}_{(I=1)}+\underbrace{(d-1)\left(\l^2+((d-2)/2)^2\right)}_{(I=2,\cdots,d)}\right]\label{trzerotv2t}\\
&=&\frac{{\cal V}_{d+1}(d)}{(4\pi)^{(d+1)/2}\G\left(\frac{d+1}{2}\right)}\int_{-\infty}^{\infty}\f{d\l }{\l^2+\n^2_v}\frac{|\G\left(\frac{d-2}{2}+i\l\right)|^2}{|\G\left(i\l\right)|^2}\left[\l^2+\left(d/2\right)^2\right] \label{trzerotv3t}\\
&=&\frac{\mathcal{V}^{}_{d+1} d}{\left(4\pi\right)^{(d+1)/2}}\frac{\left[\nu^2_v-\left(\frac{d}{2}\right)^2\right]\Gamma\left(\frac{1}{2}-\frac{d}{2}\right)\Gamma\left(\frac{d-2}{2}\pm \nu_v\right)}{\Gamma\left(\nu_v-\frac{d}{2}\right)\Gamma\left(1-\nu_v+\frac{d}{2}\right)}.\label{trzerotv4t}
\eeqa

In the above expressions $\n^2_v=M_v^2+((d-2)/2)^2$ and $\G(\pm a)=\G(a)\G(-a)$. To get to (\ref{trzerotv2t}) we have performed the $k$-integral in (\ref{trzerotv1t}). In (\ref{trzerotv2t}) we have separated out the contributions from $I=1$ and $I=2,\cdots,d$. This latter contribution is same as a set of $(d-1)$ scalars which is evident from the structure of the transverse vectors. 
The integrand in (\ref{trzerotv3t}) includes the spectral function for vectors that was previously obtained in \cite{Camporesi:1994ga}\footnote{The spectral function in equation (2.107) of \cite{Camporesi:1994ga} is denoted as $\m(\l)$. We have used the same symbol here as a measure for $\l$ integral and is defined in (\ref{mudef}).}.
In the final line (\ref{trzerotv4t}) we have performed the integral over $\lambda$ by closing the contour in the complex upper half plane (see appendix \ref{lintegralv} for details). For $d=2$ the trace matches the expression obtained for AdS$_3$ in \cite{Giombi:2008vd} using heat kernel method.

The contribution to the action for the pure-gauge longitudinal vector essentially comes from the gauge fixing and the mass terms in (\ref{scalarqedL1}). This longitudinal action for $\xi=1$ can be written as 

\beqa\label{alongv}
S_{gL}&=&-\frac{1}{2}\int d^{d+1}x\sqrt{g} \left[\tilde{A}_y^L{\cal O}\tilde{A}_y^L +\tilde{A}_y^L(y^2\pa_y+y)\pa_i\tilde{A}_i^L+\tilde{A}_i^L y^2\pa_i\pa_j\tilde{A}_j^L+\tilde{A}_i^L(y^2\pa_y-dy)\pa_i\tilde{A}_y^L-M_v^2\tilde{A}_{\m}^L\tilde{A}_{\m}^L\right]\non
&=&\frac{1}{2}\int d^{d+1}x\sqrt{g} \left[-\tilde{A}_{\m}^L{\cal O}_{\m\n}^L\tilde{A}_{\n}^L+M_v^2\tilde{A}_{\m}^L\tilde{A}_{\m}^L\right].
\eeqa

The operator ${\cal O}$ is defined in (\ref{sgdef}). The longitudinal vectors which solve the following eigenvalue equation

\beqa
{\cal O}_{\m\n}^L\tilde{A}_{\n}^L=\left[\l^2+\left(\f{d}{2}\right)^2\right]\tilde{A}_{\m}^L
\eeqa

is given by

\beqa\label{longv}
\left(\begin{array}{c}\tilde{A}_y^L\\ \tilde{A}_i^L\end{array}\right)=\f{1}{\mathscr{N}}\left(\begin{array}{c}y\pa_y\\ y\pa_i
\end{array}\right)\phi_{k,\l}(\vec{x},y)=\f{1}{|\a_L|}\left(\begin{array}{c}\left[\a_L K_{i\l}(ky)-(ky)K_{i\l+1}(ky)\right]\\\frac{ik_i}{k}(ky)K_{i\l}(ky)\end{array}\right)(ky)^{d/2}e^{i\vec{k}.\vec{x}}
\eeqa

where $\a_L=(i\l+d/2)$. Denoting $[\psi^{L}_{\vec{k},\l}(\vec{x},y)]^t= \left(\begin{array}{cc}
\tilde{A}^L_y(\vec{x}, y)&\tilde{A}^L_i(\vec{x}, y)
\end{array}\right)$, like the transverse vectors,  $[\psi^{L}_{\vec{k},\l}(\vec{x},y)]$ is normalized as (appendix (\ref{normav})) 

\beqa\label{norml}
\int d^{d+1}x\sqrt{g}~[\psi^{L}_{\vec{k}^{\prime},\l^{\prime}}(\vec{x}, y)]^{\dagger}[\psi^L_{\vec{k},\l}(\vec{x}, y)]=
(2\pi)^d k^d\d^d(\vec{k}^{\prime}-\vec{k}) \d(\l^{\prime}-\l)/\m(\l).
\eeqa

The longitudinal trace following the same steps as in (\ref{trzerotv1t})-(\ref{trzerotv4t}) is 

\beqa
\tr\left[\frac{1}{-{\cal O}^L+M_v^2}\right]&=&\int d^{d+1}x\sqrt{g}\int \f{d^dk}{(2\pi)^d}\f{1}{k^d}\int_{0}^{\infty}\f{d\l ~\m(\l)}{\l^2+\n^2_s}[\psi^{L}_{\vec{k},\l}(\vec{x},y)]^{\dagger}[\psi^L_{\vec{k},\l}(\vec{x},y)]\label{trzerotv1l}\\
&=& \frac{{\cal V}_{d+1}}{(4\pi)^{(d+1)/2}\G\left(\frac{d+1}{2}\right)}\int_{-\infty}^{\infty}\f{d\l }{\l^2+\n^2_s}\frac{|\G\left(\frac{d}{2}+i\l\right)|^2}{|\G\left(i\l\right)|^2}\label{trzerotv2l}\\
&=&\frac{\mathcal{V}^{}_{d+1} }{\left(4\pi\right)^{d+1}}\frac{\Gamma\left(\frac{1}{2}-\frac{d}{2}\right)\Gamma\left(\frac{d}{2}+ \nu_s\right)}{\Gamma\left(1-\frac{d}{2}+\n_s\right)}\label{trzerotv3l}
\eeqa

where $\n^2_s=M_v^2+(d/2)^2$. The answer is same as that of a scalar of mass $M_v$ as expected.

\subsection{Thermal AdS$_3$}\label{thermal3v}

In this section we derive the one-loop partition function for vectors on thermal AdS$_3$ which is the quotient space ${\mathbb H}^3/{\mathbb Z}$. We first illustrate the details of the computation for three dimensions and then generalize to ${\mathbb H}^{d+1}/{\mathbb Z}$ in the following section. We begin by rewriting the AdS$_3$ metric as

\beqa
ds^2=\f{L^2}{y^2}\left(dy^2+dzd\bar{z}\right)
\eeqa

with the action of $\g^{n}$ on the coordinates as,

\beqa
\g^n(y,z)=(e^{-n\beta}y, e^{2\pi in\tau}z) ~~~~\mbox{where}~~~~\tau=\frac{1}{2\pi}(\theta+i\beta).
\eeqa

In terms of real coordinates $z=x_1+ix_2$, the action of the group element on the real coordinates $\vec{x}=(x_1,x_2)$ can be written as, 

\beqa\label{gxv}
\g^n \vec{x}=\left(e^{-n\beta}(x_1 \cos n\theta-x_2\sin n\theta),e^{-n\beta}(x_1 \sin n\theta+x_2\cos n\theta)\right).
\eeqa

Defining $\mathsf{k}=k_1+ik_2$, we have $\g^n \mathsf{k}=e^{-2\pi in\bar{\tau}}\mathsf{k}$. The action of $\g^{n}$ on $\vec{k}$ is thus,

\beqa\label{gkv}
\g^n \vec{k}=\left(e^{-n\beta}(k_1 \cos n\theta+k_2\sin n\theta),e^{-n\beta}(k_2 \cos n\theta-k_1\sin n\theta)\right).
\eeqa

The action of $\g^{n}$ on the gauge fields $(A^1,A^2)$ is same as that on the coordinates (\ref{gxv}). Now let us define $\mathsf{A}^T=\tilde{A}_1^T+i\tilde{A}_2^T$ so that $\g^n \mathsf{A}^T(y,z)=e^{in\th}\mathsf{A}^T(\g^n(y,z))$. In this notation, the transverse eigenvectors are

\beqa\label{complnv}
[\psi^{T}_{\vec{k},\l}(\vec{x},y)]_{(I)}=\left(\begin{array}{c}\tilde{A}_y^T(y,z)\\ \mathsf{A}^T(y,z)\end{array}\right)_{(I)}.
\eeqa

Next define

\beqa\label{trans1t}
[\Psi^{T}_{\vec{k},\l}(\vec{x},y)]_{(I)}=\f{1}{{\cal N}}\sum_{n=-\infty}^{\infty} \g^n[\psi^{T}_{\vec{k},\l}(\vec{x},y)]_{(I)}= \f{1}{{\cal N}}\sum_{n=-\infty}^{\infty} \left(\begin{array}{c}\tilde{A}_y^T(\g^n(y,z))\\ e^{i n\th}\mathsf{A}^T(\g^n(y,z))\end{array}\right)_{(I)}.
\eeqa 

The function $\Psi^{T}_{\vec{k},\l}(\vec{x},y)$ is invariant under thermal identification allowing us to integrate over the full $\mathbb{H}^3$ instead of the fundamental domain of $\mathbb{H}^3/\mathbb{Z}$. This simplification leads to an infinite over counting
which we take care of by introducing the normalization ${\cal N}$. An identical setup can be followed for the longitudinal vector.   

In the following part of this section we show the details of the computations for the transverse $I=1$ vector using (\ref{trans1}) and (\ref{trans1t}). The related calculations for $I=2$ and the longitudinal vector being similar are shown in appendix \ref{acomptv}. We first work out the normalization for these functions for $I=1$ in terms of $\vec{k}$ and $\l$. For this consider the following integral 

\beqa
&&\int d^{3}x~\sqrt{g}~[\Psi^{T}_{\vec{k^{\prime}},\l^{\prime}}(\vec{x},y)]^{\dagger}_{(1)}[\Psi^T_{\vec{k},\l}(\vec{x},y)]_{(1)}\non
&=&\f{1}{{\cal N}^2}\sum_{n,n^{\prime}}\int d^{3}x~\sqrt{g}\left[\tilde{A}^{T*}_y(\g^{n^{\prime}}(y,z))\tilde{A}^{T}_y(\g^n(y,z))+e^{i(n-n^{\prime})\th}\bar{\mathsf{A}}^T(\g^{n^{\prime}}(y,z))\mathsf{A}^T(\g^n(y,z))\right]\\
&=&\f{1}{{\cal N}^2}\sum_{n,n^{\prime}}\int\f{dy}{y}(ke^{-n\b})(k^{\prime}e^{-n^{\prime}\b})\left[(ke^{-n\b} y)(k^{\prime}e^{-n^{\prime}\b}y)K_{-i\l^{\prime}}(k^{\prime}e^{-n^{\prime}\b}y)K_{i\l}(ke^{-n\b}y)\right.\label{normI1v1}\non
&+&\left.e^{i(n-n^{\prime})\th}\f{\bar{\mathsf{k}}^{\prime}\mathsf{k}}{k^{\prime}k}\f{1}{|\a_T(\l^{\prime})|}\left\{\a_T^*(\l^{\prime})K_{-i\l^{\prime}}(k^{\prime}ye^{-n^{\prime\b}})-(k^{\prime}ye^{-n^{\prime}\b})K_{-i\l^{\prime}+1}(k^{\prime}ye^{-n^{\prime}\b})\right\}\right.\non
&\times&\left. \f{1}{|\a_T(\l)|}\left\{\a_T(\l)K_{i\l}(kye^{-n\b})-(kye^{-n\b})K_{i\l+1}(kye^{-n\b})\right\}\right](2\pi)^2\d^2(\g^n\vec{k}-\g^{n^{\prime}}\vec{k}^{\prime})\label{normI1v2}\\
&=& \f{1}{{\cal N}}\sum_{n}(ke^{-n\b})^2(2\pi)^2\d^2(\g^{n} \vec{k}-\vec{k}^{\prime})\d(\l-\l^{\prime})/\m(\l)\label{normI1v3}\\&=&(2\pi)^2k^2\d^2(\vec{k}-\vec{k}^{\prime})\d(\l-\l^{\prime})/\m(\l)\label{normI1v4}.
\eeqa

The normalization works out to be same as that of a scalar field with $\m(\l)$ defined in (\ref{mudef}). Several simplifications have been made in going from (\ref{normI1v1}) to (\ref{normI1v3}). In (\ref{normI1v2}) we have performed the integral over $\vec{x}$ which leads to a delta function. Imposing this delta function and using the identity $\d^2(\g^n\vec{k}-\g^{n^{\prime}}\vec{k}^{\prime})=e^{2n^{\prime}\b}\d^2(\g^{n-n^{\prime}}\vec{k}-\vec{k}^{\prime})$, the resulting expression contains $(n,n^{\prime})$ dependent terms of the form $(n-n^{\prime})$. Due to this symmetry, the summation over $n^{\prime}$ has been done canceling a factor of ${\cal N}$ in the denominator. Further noting that  $e^{i(n-n^{\prime})\th}{\bar{\mathsf{k}}^{\prime}\mathsf{k}}/{k^{\prime}k}=1$ (because of the delta function) the resulting integral over $y$ can be performed as in the zero temperature case as shown in appendix \ref{normav}. The final expression is obtained by absorbing an infinite coefficient resulting (from the sum over $n$) into the remaining ${\cal N}$.

The trace (\ref{trv}) can now be computed for $I=1$ using (\ref{trans1}) and (\ref{trans1t}).

\beqa\label{tracevt}
&&\tr\left[\frac{1}{-{\cal O}^T+M_v^2-d+1}\right]_{(1)}=
\int d^{3}x\sqrt{g}\int \f{d^2k}{(2\pi)^2}\f{1}{k^2}\int_{0}^{\infty}\f{d\l ~\m(\l)}{\l^2+\n^2_v}[\Psi^{T}_{\vec{k},\l}(\vec{x},y)]^{\dagger}_{(1)}[\Psi^T_{\vec{k},\l}(\vec{x},y)]_{(1)}\non
&=&
\f{1}{{\cal N}^2}\sum_{n,n^{\prime}}\int \f{dy}{y}\int\f{d^2k}{(2\pi)^2}\int\f{d\l ~\m(\l)}{\l^2+\n^2_v}\f{1}{|\a_T|^2}e^{-(n+n^{\prime})\b}
\left[|\a_T|^2e^{i(n-n^{\prime})\th} K_{i\l}(kye^{-n\b})K_{-i\l}(kye^{-n^{\prime}\b})\right.\non
&+&\left.(kye^{-n\b})(kye^{-n^{\prime}\b})\left\{K_{i\l}(kye^{-n\b})K_{-i\l}(kye^{-n^{\prime}\b})
+e^{i(n-n^{\prime})\th}K_{i\l+1}(kye^{-n\b})K_{-i\l+1}(kye^{-n^{\prime}\b})\right\}\right.\non
&-&\left.e^{i(n-n^{\prime})\th}(ky)\left\{\a_T e^{-n^{\prime}\b}K_{i\l}(kye^{-n\b})K_{-i\l+1}(kye^{-n^{\prime}\b})+\a_T^*e^{-n\b}K_{-i\l}(kye^{-n^{\prime}\b})K_{i\l+1}(kye^{-n\b})\right\}
  \right]\non
  &\times& (2\pi)^2\d^2(\g^{n}\vec{k}-\g^{n^{\prime}}\vec{k}).
  \eeqa

The above expression is obtained after performing the integral over $\vec{x}$ which gives the momentum delta function. We now follow the steps in \cite{Kakkar:2022hub,Kakkar:2023gzu} to simplify the above expression which include using the following identity

\beqa\label{di2v}
\d^2(\g^n\vec{k}-\g^{n^{\prime}}\vec{k})=\frac{e^{2n^{\prime}\b}}{|1-e^{2\pi i (n-n^{\prime})\t}|^2}\d^2(\vec{k}).  
\eeqa

We next scale $y \rightarrow y e^{n^{\prime}\b}$. In the resulting expression the sum over $n,n^{\prime}$ can be converted into a single sum over $n$ and cancelling a factor of normalization ${\cal N}$. We discard the $n=0$ term which corresponds to the zero temperature piece (\ref{trzerotv4t}) with $d=2$. Further noting the invariance of the expression for $n \rightarrow -n$ and the integrand for $\l \rightarrow -\l$ we can write down the following expression for the trace

\beqa  
&&\f{1}{{\cal N}}\sum_{n=1}\f{e^{-n\b}}{|1-e^{2\pi in\t}|^2}\int\f{dy}{y}\int\f{d^2k}{(2\pi)^2}\int_{-\infty}^{\infty}\f{d\l ~\m(\l)}{\l^2+\n^2_v}\f{1}{|\a_T|^2}
\left[|\a_T|^2\cos(n\th)K_{i\l}(kye^{-n\b})K_{-i\l}(ky)\right.\non
&+&\left.(kye^{-n\b})(ky)\left\{K_{i\l}(kye^{-n\b)}K_{-i\l}(ky)
+\cos(n\th)K_{i\l+1}(kye^{-n\b})K_{-i\l+1}(ky)\right\}\right.\non
&-&\left.\cos(n\th)(ky)\left\{\a_T K_{i\l}(kye^{-n\b})K_{-i\l+1}(ky)+\a_T^*e^{-n\b}K_{-i\l}(ky)K_{i\l+1}(kye^{-n\b})\right\}
  \right]\non
  &\times& (2\pi)^2\d^2(k).
\eeqa

Note that the first term in the integrand is same as a scalar (except for the factor of $\cos(n\th)$) and was worked out in \cite{Kakkar:2022hub}. Thus

\beqa\label{tracevtt1}
&&\f{1}{{\cal N}}\sum_{n \ne 0}\f{e^{-n\b}}{|1-e^{2\pi i n\t}|^2}\int\f{dy}{y}\int\f{d^2k}{(2\pi)^2}\int\f{d\l ~\m(\l)}{\l^2+\n^2_v}
\left[\cos(n\th)K_{i\l}(kye^{-n\b})K_{-i\l}(ky)\right](2\pi)^2\d^2(k)\non
&=&\sum_{n=1}^{\infty}\cos(n\th)\f{\b e^{-n\b(1+\n_v)}}{\n_v|1-e^{2\pi i n\t}|^2}.
\eeqa

In the last line we have performed the $y$-integral by regularizing it as follows so that only the fundamental domain $e^{-\b}\le y \le 1$ contributes to the trace 

\beqa
\int_0^{\infty} \f{dy}{y}=\sum_{m=-\infty}^{\infty}\int_{e^{-(m+1)\b}}^{e^{-m\b}} \f{dy}{y}
={\cal N}\b~.
\eeqa

The contribution from the other terms add up to zero. To see this, we look at these terms in the $k\rightarrow 0$ limit as imposed by the delta function

\beqa\label{limitsv}
&&(kye^{-n\b})(ky)K_{i\l}(kye^{-n\b})K_{-i\l}(ky) \xrightarrow[\text{}]{\text{{$k\rightarrow 0$}}} 0\\
&&(ky)K_{i\l}(kye^{-n\b})K_{-i\l+1}(ky)
\xrightarrow[\text{}]{\text{{$k\rightarrow 0$}}} \f{1}{2}e^{in\b\l}\G(1-i\l)\G(i\l)\non
&&(kye^{-n\b})K_{-i\l}(ky)K_{i\l+1}(kye^{-n\b})
\xrightarrow[\text{}]{\text{{$k\rightarrow 0$}}} \f{1}{2}e^{in\b\l}\G(1+i\l)\G(-i\l)\non
&&(kye^{-n\b})(ky)K_{i\l+1}(kye^{-n\b})K_{-i\l+1}(ky) \xrightarrow[\text{}]{\text{{$k\rightarrow 0$}}} e^{in\b\l}|\G(1+i\l)|^2.\nonumber
\eeqa

The integral over $\lambda$ with $\m(\l)$ in (\ref{mudef}) is then

\beqa\label{othertv}
\int_{-\infty}^{\infty}\f{d\l ~\m(\l)}{\l^2+\n^2_v}\f{e^{in\b\l}}{|\a_T|^2}\left[-\f{1}{2}\left\{ (-i\l)\a_T+(i\l)\a_T^*\right\}+|i\l|^2\right]|\G(i\l)|^2=0
\eeqa

as the integrand vanishes. For the other transverse vector (\ref{trans23}) we get the same expression as (\ref{tracevtt1}). See appendix \ref{acomptv} for details.
The finite temperature transverse partition function can then be written as 

\beqa\label{transtt3}
\log Z^T_{\t}&=&\f{1}{2}\int_{M_v^2}^{\infty}\tr\left[\frac{1}{-{\cal O}^T+M_v^2-d+1}\right]dM_v^2\\
&=&\sum_{n=1}^{\infty}2\cos(n\th)\f{\b e^{-n\b}}{|1-e^{2\pi i n\t}|^2}\int_{M_v^2}^{\infty} \f{e^{-n\b\n_v}}{2\n_v} dM_v^2\\
&=&\sum_{n=1}^{\infty}2\cos(n\th)\f{1}{n}\f{e^{-n\b\D_v}}{|1-e^{2\pi i n\t}|^2}=\sum_{n=1}^{\infty}\f{1}{n}\f{q^n+\bar{q}^n}{|1-q^n|^2}|q|^{n M_v}
\eeqa

where $\D_v=1+\n_v=1+|M_v|$ and $q=e^{2\pi i\t}$. For $M_v=0$ the expression matches the one obtained in \cite{Giombi:2008vd}, \cite{David:2009xg} using heat kernel method. The above expression can also be rewritten in the following form that makes the connection with the boundary CFT partition function transparent, 

\beqa\label{transtt31}
\log Z^T_{\t}&=&\sum_{n=1}^{\infty}\sum_{l,l^{\prime}=0}^{\infty}\f{1}{n}\left[q^{n(1+l+M_v/2)}\bar{q}^{n(l^{\prime}+M_v/2)}+q^{n(l+M_v/2)}\bar{q}^{n(1+l^{\prime}+M_v/2)}\right]\\
&=&-\sum_{l,l^{\prime}=0}^{\infty}\left[\log\left(1-q^{(1+l+M_v/2)}\bar{q}^{(l^{\prime}+M_v/2)} \right)+\log\left(1-q^{(l+M_v/2)}\bar{q}^{(1+l^{\prime}+M_v/2)} \right)\right]\\
&=& -2\sum_{l=0}^{\infty}(l+1)\log\left(1-e^{-\b(\D_v+l)}\right)~~~~~~~\mbox{for $\th=0$}.
\eeqa

Some details of the computation for the longitudinal vector are shown in appendix \ref{acomptv}. 
This contribution to the finite temperature partition function is same as a scalar of mass $M_v$

\beqa\label{longtt3}
\log Z^L_{\t}&=&\sum_{n=1}^{\infty}\f{\b e^{-n\b}}{|1-e^{2\pi i n\t}|^2}\int_{M_v^2}^{\infty} \f{e^{-n\b\n_s}}{2\n_s} dM_v^2\\
&=&\sum_{n=1}^{\infty}\f{1}{n}\f{e^{-n\b\D_s}}{|1-e^{2\pi i n\t}|^2}=\sum_{n=1}^{\infty}\f{1}{n}\f{|q|^{n\D_s}}{|1-q^n|^2}\\
&=&-\sum_{l=0}^{\infty}(l+1)\log\left(1-e^{-\b(\D_s+l)}\right)~~~~~~~\mbox{for $\th=0$}.
\eeqa
  
where $\D_s=1+\n_s=1+\sqrt{1+M_v^2}$.

\subsection{Thermal AdS$_{d+1}$}\label{thermaldv}

The generalization of results in the previous section to $d+1$ dimensions follows along the same lines as in \cite{Kakkar:2022hub}, \cite{Kakkar:2023gzu}. Some details of the computation are however different which we highlight here. To begin with the basic setup,  let us first consider the case when $d$ is even. The metric is then written as
 
\beqa
ds^2=\f{L^2}{y^2}\left(dy^2+\sum_{i=1}^{d/2} dz_id\bar{z_i}\right).
\eeqa

The thermal AdS$_{d+1}$ space we consider has the conformal boundary $S^1\times S^{d-1}$. This space is the quotient space $\mathbb{H}^{d+1}/\mathbb{Z}$. The action of $\gamma^n \in \mathbb{Z}$ on the Poincar\'e coordinates is,

\beqa\label{realtdv}
\g^n(y, z_1, \cdots, z_{d/2})=(e^{-n\b}y,e^{2\pi i n\t_1}z_1,\cdots, e^{2\pi i n\t_{d/2}}z_{d/2}).
\eeqa

Next define real coordinates $z_i=x_{i1}+ix_{i2}$. The action of $\g^n$ on the real coordinates $\vec{x}_i=(x_{i1},x_{i2})$ 
is given by equation (\ref{gxv}) with $\theta$ replaced by $\theta_i$. In the similar way we define $\mathsf{A}_i=\tilde{A}_{2i-1}+i\tilde{A}_{2i}$ so that $\g^{n}\mathsf{A}_i=e^{in\th_i}\mathsf{A}_i$ and $\mathsf{k}_i=k_{2i-1}+ik_{2i}$ with $\g^{n}\mathsf{k}_i=e^{-2\pi in\bar{\t}_i}\mathsf{k}_i$.
We can now write down a $d/2+1$ component transverse vactor as in (\ref{complnv}) with the corresponding action of $\g$ as defined in (\ref{trans1t}). Let us consider the case for AdS$_5$, the results can then be easily generalized for arbitrary even $d$. The normalization works out exactly like the AdS$_3$ so we do not repeat it here. We proceed to compute the trace. Apart from the transverse vector (\ref{trans1}), the other three transverse vectors are

\beqa\label{ads5t2}
[\psi^{T}_{\vec{k},\l}(\vec{x},y)]^t_{(2)}=\left(\begin{array}{ccc}0,&-i\mathsf{k}_1/|\mathsf{k}_1|,&0\end{array}\right)\phi_{k,\l}(\vec{x},y)
\eeqa
\vspace{-1cm}
\beqa\label{ads5t3}
[\psi^{T}_{\vec{k},\l}(\vec{x},y)]^t_{(3)}=\left(\begin{array}{ccc}0,& 0,&-i\mathsf{k}_2/|\mathsf{k}_2|\end{array}\right)\phi_{k,\l}(\vec{x},y)
\eeqa
\vspace{-1cm}
\beqa\label{ads5t4}
[\psi^{T}_{\vec{k},\l}(\vec{x},y)]^t_{(4)}=\left(\begin{array}{ccc}0,&\mathsf{k}_1|\mathsf{k}_2|/(|\mathsf{k}_1|k),&\mathsf{k}_2|\mathsf{k}_1|/(|\mathsf{k}_2|k)\end{array}\right)\phi_{k,\l}(\vec{x},y).
\eeqa

Using these we can now write down the invariant functions $[\Psi^{T}_{\vec{k},\l}(\vec{x},y)]_{(I)}$ as in (\ref{trans1t}).
  The contribution to the trace from $[\Psi^{T}_{\vec{k},\l}(\vec{x},y)]_{(1)}$ following previous computational steps in (\ref{tracevt}) is

\beqa\label{trevendv1}  
&&\f{1}{{\cal N}}\sum_{n=1}^{\infty}\int\f{dy}{y}\int\f{d^4k}{(2\pi)^4}\int_{-\infty}^{\infty}\f{d\l ~\m(\l)}{\l^2+\n^2_v}\f{1}{|\a_T|^2}
\left[|\a_T|^2f_{(1)}(\th_1,\th_2)K_{i\l}(kye^{-n\b})K_{-i\l}(ky)\right.\non
&+&\left.(kye^{-n\b})(ky)\left\{K_{i\l}(kye^{-n\b)}K_{-i\l}(ky)
+f_{(1)}(\th_1,\th_2)K_{i\l+1}(kye^{-n\b})K_{-i\l+1}(ky)\right\}\right.\non
&-&\left.f_{(1)}(\th_1,\th_2)(ky)\left\{\a_T K_{i\l}(kye^{-n\b})K_{-i\l+1}(ky)+\a_T^*e^{-n\b}K_{-i\l}(ky)K_{i\l+1}(kye^{-n\b})\right\}
  \right]\non
  &\times& \prod_{i=1}^2\f{e^{-n\b}}{|1-e^{2\pi in\t_i}|^2}(2\pi)^2\d^2(k_i)
\eeqa

\beqa
f_{(1)}(\th_1,\th_2)=\left(|\mathsf{k}_1|^2\cos (n\th_1)+|\mathsf{k}_2|^2\cos (n\th_2)\right)/k^2.
\eeqa

Due to (\ref{othertv}), only the first term in the expression (\ref{trevendv1}) survives the $k \rightarrow 0$ limit imposed by the momentum delta function. The contribution to the trace from the other three transverse vectors (\ref{ads5t2})-(\ref{ads5t4}) is

\beqa\label{trevendv2}  
&&\f{1}{{\cal N}}\sum_{n=1}^{\infty}\int\f{dy}{y}\int\f{d^4k}{(2\pi)^4}\int_{-\infty}^{\infty}\f{d\l ~\m(\l)}{\l^2+\n^2_v}
K_{i\l}(kye^{-n\b})K_{-i\l}(ky)\non
&\times& \left[f_{(4)}(\th_1,\th_2)+\cos(n\th_1)+\cos(n\th_2)\right]\prod_{i=1}^2\f{e^{-n\b}}{|1-e^{2\pi in\t_i}|^2}(2\pi)^2\d^2(k_i)
\eeqa 

where $f_{(4)}(\th_1,\th_2)$ comes from (\ref{ads5t4}) and is given by

\beqa
f_{(4)}(\th_1,\th_2)=\left(|\mathsf{k}_2|^2\cos (n\th_1)+|\mathsf{k}_1|^2\cos (n\th_2)\right)/k^2.
\eeqa

Adding (\ref{trevendv1}) and (\ref{trevendv2}) we thus have for $d=4$ and for general even $d$ 

\beqa\label{partads4ve}
\log Z_{\t}^T&=& \sum_{n=1}^{\infty}2\left[\cos(n\th_1)+\cos(n\th_2)\right]\f{e^{-n\b\n_v}}{n}\prod_{i=1}^2\f{e^{-n\b}}{|1-e^{2\pi in\t_i}|^2}\\
&\rightarrow&\sum_{n=1}^{\infty}2\left[\cos(n\th_1)+\cdots+\cos(n\th_{d/2})\right]\f{e^{-n\b\n_v}}{n}\prod_{i=1}^{d/2}\f{e^{-n\b}}{|1-e^{2\pi in\t_i}|^2}\\
&=& \sum_{n=1}^{\infty}\sum_{k=1}^{d/2}(q^n_k+\bar{q}^n_k)\prod^{d/2}_{i=1}\sum_{l_i,l_i^{\prime}=0}^{\infty}\f{1}{n}q_i^{n(l_i+(\D_v-1)/d)}\bar{q}_i^{n(l_i^{\prime}+(\D_v-1)/d)}\\
&=&-\sum_{k=1}^{d/2}\sum_{\substack{l_1,l^{\prime}_1\\...\\l_{d/2}l^{\prime}_{d/2}=0}}^{\infty}\left[\log\left(1-q_k\prod^{d/2}_{i=1}q_i^{l_i+(\D_v-1)/d}\bar{q}_i^{l_i^{\prime}+(\D_v-1)/d}\right)+(q_k\rightarrow \bar{q}_k) \right]\non
&=&-d\sum_{\substack{l_1,l^{\prime}_1\\...\\l_{d/2}l^{\prime}_{d/2}=0}}^{\infty}\log\left(1-e^{-\b(l_1+l_1^{\prime}+\cdots+l_{d/2}+l_{d/2}^{\prime} +\D_v)}\right) ~~~~\mbox{for $\th_i=0$}\\
&=&-d \sum_{l=0}^{\infty}g(l,d)\log\left(1-e^{-\b(\D_v+l)}\right)~~~~\mbox{where~~ $g(l,d)=\f{(d+l-1)!}{l!(d-1)!}$}.\label{degenv}
\eeqa

In the above expressions  $\Delta_v=\frac{1}{2}\left(d+ \sqrt{(d-2)^2+4M_v^2}\right)$. The expression on the r.h.s. of (\ref{partads4ve}) for $M_v=0$ for AdS$_5$ matches with the result in \cite{Gopakumar:2011qs,Gupta:2012he} where the partition function for spin $s$ was derived in terms of $SO(d)$ characters.

For odd $d$ the metric is written as

\beqa
ds^2=\f{L^2}{y^2}\left(dy^2+\sum_{i=1}^{(d-1)/2} dz_id\bar{z_i}+dx^{2}_{d}\right)
\eeqa 

with the following action of $\g^{n}$ on the coordinates,

\beqa
\g^n(y, z_1, \cdots z_{(d-1)/2},x_d)=(e^{-n\b}y,e^{2\pi i n\t_1}z_1,\cdots, e^{2\pi i n\t_{(d-1)/2}}z_{(d-1)/2},e^{-n\b}x_d).
\eeqa

Let us write as before $\mathsf{A}_i=\tilde{A}_{2i-1}+i\tilde{A}_{2i}$ so that $\g^{n}\mathsf{A}_i=e^{in\th_i}\mathsf{A}_i$ and $\mathsf{k}_i=k_{2i-1}+ik_{2i}$ with $\g^{n}\mathsf{k}_i=e^{-2\pi in\bar{\t}_i}\mathsf{k}_i$. Some essential features of the computation for AdS$_4$ are given in appendix \ref{acomptv}. Then following similar computation for general odd $d$,

\beqa
\log Z_{\t}^T
&=&\sum_{n=1}^{\infty}\left[2\left(\cos(n\th_1)+\cdots+\cos(n\th_{(d-1)/2})\right)+1\right]\f{e^{-n\b(\n_v+1/2)}}{n(1-e^{-n\b})}\prod_{i=1}^{(d-1)/2}\f{e^{-n\b}}{|1-e^{2\pi in\t_i}|^2}\\
&=&-\sum_{k=1}^{(d-1)/2}\sum_{\substack{l_1,l^{\prime}_1\\...\\l_d=0}}^{\infty}\left[\log\left(1-q_k|q_d|^{l_d}\prod^{(d-1)/2}_{i=1}q_i^{l_i+(\D_v-1)/(d-1)}\bar{q}_i^{l_i^{\prime}+(\D_v-1)/(d-1)}\right)+(q_k\rightarrow \bar{q}_k) \right]\non
&+& \sum_{\substack{l_1,l^{\prime}_1\\...\\l_d=0}}^{\infty}\log\left(1-|q_d|^{l_d}\prod^{(d-1)/2}_{i=1}q_i^{l_i+(\D_v-1)/(d-1)}\bar{q}_i^{l_i^{\prime}+(\D_v-1)/(d-1)}\right)\\
&=&-d\sum_{\substack{l_1,l^{\prime}_1\\...\\l_d=0}}^{\infty}\log\left(1-e^{-\b(l_1+l_1^{\prime}+\cdots+l_{(d-1)/2}+l_{(d-1)/2}^{\prime}+l_d +\D_v)}\right) ~~~~\mbox{for $\th_i=0$}\\
&=& -d\sum_{l=0}^{\infty}g(l,d)\log\left(1-e^{-\b(\D_v+l)}\right) 
\eeqa

where $\D_v$ is defined below (\ref{partads4ve}) and $g(l,d)$ is given in (\ref{degenv}). The contribution to the partition function from the longitudinal vector which is same as a scalar, is computed exactly as shown in (\ref{longvt}) for AdS$_3$ for both even and odd $d$. For $\th_i=0$ the finite temperature expression for the one-loop gauge partition function including the longitudinal vector and the ghost is

\beqa\label{finittegauge}
\log Z_{\tiny{\mbox{gauge}}}+\log Z_{\tiny{\mbox{ghost}}}&=&\sum_{n=1}^{\infty}\frac{1}{n}\frac{d e^{-\Delta_v n\b}-e^{-\Delta_s n\b}}{\left(1-e^{-n\b}\right)^d}\non
&=&\sum_{l=0}^{\infty}g(l,d)\left[\log\left(1-e^{-\b(\D_s+l)}\right)-d\log\left(1-e^{-\b(\D_v+l)}\right)\right]
\eeqa

where, $\D_v$ is given below (\ref{partads4ve}) and $\Delta_s=\frac{1}{2}\left(d+ \sqrt{d^2+4M_v^2}\right)$. The above result has been obtained previously using other methods in \cite{Gibbons:2006ij}, \cite{Gopakumar:2011qs}, \cite{Gupta:2012he}.  Along with the integrated zero temperature traces (\ref{trzerotv4t}), (\ref{trzerotv3l}) and ghost, we shall use (\ref{finittegauge}) in the following sections for analysis of phases of scalar QED theories.

\section{Effective potentials and phases of scalar QED theories}\label{phasesv}

In this section we study the phases of the single-scalar QED theory and the theory for $N$ scalar fields for $d=2,3$. The corresponding effective potentials needed for these studies can be written down from results for vector partition functions in the previous sections and scalar trace from \cite{Kakkar:2022hub}. We have seen in all the computations that the zero temperature contribution to the trace for all fields is proportional to the divergent volume of Euclidean AdS space. However for the finite temperature contributions this volume does not factor out. We thus need to regulate this divergence. We shall use the following regularized volumes in our computations: ${\cal V}_3=-\pi\b/2$, ${\cal V}_4=2\pi\b/3$. The computation of these is reviewed in \cite{Kakkar:2022hub}. 

In the following sections we shall first study the phases as a function of the AdS radius $L$ at zero temperature. For the analysis at finite temperature, we shall also fix $L=1$ and set $\th_i=0$. The phases are studied as a function of $\b$ and the scalar mass-squared, $m^2$. Putting back factors of $L$ at finite temperature, one can study a richer phase structure of these theories as a function of the AdS radius for finite temperature also. We however do not pursue it here. In the numerical analysis the sum over the finite summations would be truncated at $n_{\mbox{\tiny{max}}}=10$ as the series converges to the desired accuracy within a first few terms. 

\subsection{Single-scalar QED}\label{phasesssqed}

The general form of the effective potential from (\ref{potsqed}) is given by

\beqa
\label{potsqedtr}
V^{}_{eff}(\phi^{}_{cl})&=&
\f{1}{{\cal V}_{d+1}}\int_0^{M_v^2}\left\{\f{1}{2}\mbox{tr}\left[\f{1}{-\mathcal{O}^T+M_v^2-d+1}\right]+\f{1}{2}\mbox{tr}\left[\f{1}{-\mathcal{O}^L+ \xi M_v^2}\right]
-\mbox{tr}\left[\f{1}{-\square^{}_{E}+\xi M_v^2}\right]\right\}dM_v^2\non
&+&\frac{1}{2{\cal V}_{d+1}}\sum_{i=1,2}\int_0^{M_i^2}\mbox{tr}\left[\f{1}{-\square^{}_{E}+M_i^2}\right]dM_i^2+m^2\phi_{cl}^2+\f{\l}{4}\phi_{cl}^4+\mbox{counterterms}.
\eeqa
   
In the above expression the first two terms inside the integral are contributions from the transverse and longitudinal vectors. The third term is the contribution from the ghost. The sum over $i$ in the fourth term denotes contributions from the scalars. The counterterms in (\ref{potsqedtr}) absorb the UV divergences which arise from the zero-temperature contributions. These divergences can be removed using standard renormalization procedures as in flat space, for example, using the minimal subtraction scheme or by implementing renormalized perturbation theory. The dimensionally regularised expressions for the traces have poles only for odd $d$. For even $d$ the expressions give finite expressions. We shall thus impose renormalization conditions only for AdS$_4$ where divergences arising from two-point and from four-point functions are absorbed into counterterms for scalar mass $m^2$ and the coupling constant $\l$. The UV divergence from the QED vertex is absorbed into the counterterm for $e$. Since this is not an essential part of our one-loop analysis, in the expressions for the effective potentials, it will be assumed that the charge $e$ that appears is the renormalized charge. For AdS$_3$ since the expression is already finite, we shall assume that $m^2$ is the renormalized mass. This is similar to  using the minimal subtraction scheme for AdS$_3$. We note that implementing the alternate scheme of imposing renormalization condition for the two point in AdS$_3$ shifts the renormalized mass. This was shown in \cite{Kakkar:2022hub}. Here however we shall see that this scheme cannot be implemented as the potential is non-differentiable at $\phi_{cl}=0$. Finally for simplicity we have set to zero the $|\phi_{cl}|^6$ term in the AdS$_3$ analysis.

\subsubsection{AdS$_3$}\label{phases3v}
The zero temperature contribution from the transverse vector trace (\ref{trzerotv4t}) is

\beqa
\mbox{tr}\left[\f{1}{-\mathcal{O}^T+M_v^2-d+1}\right]=\frac{{\cal V}_{3}}{(2\pi)}\f{1}{\sqrt{M_v^2}}(1-M_v^2).
\eeqa

It appears that the trace diverges for $M_v=0$, however this is an artifact of the computation as we have taken derivative with respect to $M_v^2$ instead of $M_v$ in order get to (\ref{potsqed}) from (\ref{trv}). The contribution to the effective potential as in (\ref{potsqedtr}) is ultimately finite. For $M_v=0$ the potential from the transverse vectors (the first term in equation (\ref{potzerot3v}) below) vanishes as was noted in 
\cite{Giombi:2008vd}.

Including the longitudinal, ghost and scalar traces we get the following effective potential at zero temperature

\begin{equation}\label{potzerot3v}
V_{eff}^0=m^2\phi_{cl}^2+\frac{\l}{4}\phi_{cl}^4+\frac{3(M_v^2)^{1/2}-(M_v^2)^{3/2}}{6 \pi}+\frac{\left(1+\xi M_v^2\right)^{3/2}}{12\pi}-\frac{\left(1+M_1^2\right)^{3/2}}{12\pi}-\frac{\left(1+M_2^2\right)^{3/2}}{12\pi}.
\end{equation}

The theory with massless scalars has three dimensionful parameters $e$, $\l$ and $L$. For the one-loop correction to be of the same order in perturbation theory as the tree-level we shall consider $\l \sim e^2$. We now put back factors of AdS radius $L$ and consider the large $L$  expansion

 \begin{eqnarray}\label{potzerot3flatv}
V_{eff}^0 &\sim &m^2\phi_{cl}^2+\frac{\l}{4}\phi_{cl}^4-\f{1}{12\pi}\left[2(M_v^2)^{3/2}-\left(\xi M_v^2\right)^{3/2}+\left(M_1^2\right)^{3/2}+\left(M_2^2\right)^{3/2}\right]\non
&+&\f{1}{8\pi L^2}\left[4(M_v^2)^{1/2}+(\xi M_v^2)^{1/2}-(M_1^2)^{1/2}-(M_2^2)^{1/2}\right]+{\cal O}(1/L^4).
\end{eqnarray}

The leading terms in the above expression give the one-loop result. This was obtained for flat space in \cite{Tan:1996kz,Tan:1997ew}\footnote{The analysis in \cite{Tan:1996kz,Tan:1997ew} also involves a two-loop computation. Here however we restrict to the one-loop analysis.}. We note that unlike flat-space, for AdS$_3$ the one-loop correction to the potential is not differentiable at $\phi_{cl}=0$ due to presence of the term linear in $|M_v|$. This is similar to that of spin-half fermions in AdS$_3$ \cite{Kakkar:2023gzu}. This similarity is also reflected in the fact that $\D_f=1+|M_f|$ for fermion and $\D_v=1+|M_v|$ have the same form. 

Let us now point out some features of the potential for massless ($m^2=0$) scalars with ($\xi=1$). $m^2$ can be negative in AdS$_3$ up to the BF bound $-1$. For negative $m^2$ there is symmetry breaking at classical level. 
Since the potential is symmetric about $\phi_{cl}=0$, we work with the right ($\phi_{cl}>0$) branch of the potential. For small values of $\phi_{cl}$ the potential is

\begin{eqnarray}\label{potzerot3s}
V_{eff}^0 &\sim & -\frac{1}{12 \pi  L^3}+\frac{\sqrt{2}e\phi_{cl}}{2\pi  L^2}-\frac{\l \phi_{cl}^2}{4\pi L}-\frac{\sqrt{2} e^3 \phi_{cl}^3}{3 \pi }+{\cal O}\left(\phi_{cl}^4\right).
\end{eqnarray}

\begin{figure}[H] 
\begin{center} 
  \begin{minipage}{0.45\textwidth}%
   \begin{subfigure}[b]{0.9\linewidth}
    \centering
    \includegraphics[width=1\linewidth]{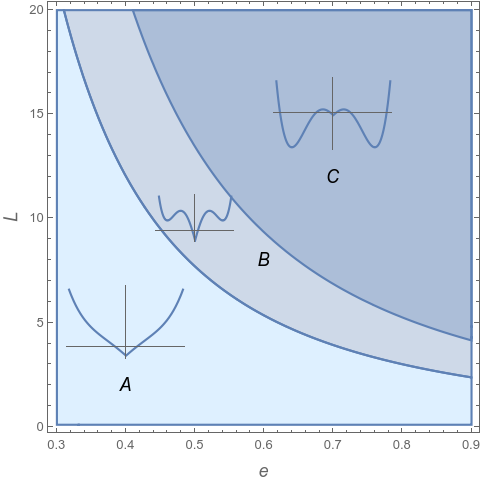} 
    \caption{} 
    \label{EAdS3} 
    \vspace{1ex}
  \end{subfigure}
  \end{minipage}
 \begin{minipage}{0.45\textwidth}%
   \begin{subfigure}[b]{0.9\linewidth}
    \centering
    \includegraphics[width=1\linewidth]{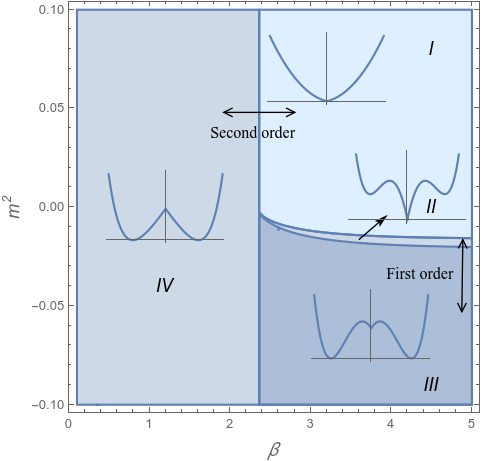} 
    \caption{} 
    \label{TAdS3} 
    \vspace{1ex}
  \end{subfigure}
\end{minipage}
  \caption{Phases as a function of (a) $L$ and $e$ at zero temperature and $m^2=0$ (b) $\b$ and $m^2$ for $e=0.1$ and $L=1$.}
  \label{VAdS3} 
  \end{center} 
\end{figure}

From the above expansion, the first derivative of the potential at $\phi_{cl}=0+$ is positive while the second and third derivatives are negative. The potential plot starts rising $\phi_{cl}=0+$ and develops a minimum at $\phi_{cl}=\phi_0$ depending on the values of $L$ and $e$ shown as region $B$ in the phase plot in figure \ref{EAdS3}. In region $C$ of the same figure, $V(\phi_0)<V(0)$. Thus there is a first order transition across regions $B$ and $C$. At large values of $L$ compared to $1/e^2$ the potential approaches that of flat space. Region $A$ has only the minimum at $\phi_{cl}=0$.

At finite temperature, collecting the expressions from (\ref{transtt3}), (\ref{longtt3}) the potential is

\beqa\label{ads3ftv}
V_{eff}=V_{eff}^0+\f{2}{\pi\b}\sum_{n=1}^{\infty}\f{1}{n(1-e^{-n\b})^2}\left[2e^{-n\b\D_v}-e^{-n\b\D_s}+e^{-n\b\D_1}+e^{-n\b\D_2}\right]
\eeqa

where 

\beqa\label{deltaads3}
\D_v=1+|M_v|~~~~~\D_s=1+\sqrt{1+M_v^2}~~~~~\D_i=1+\sqrt{1+M_i^2}
\eeqa

At finite temperature we study the phases setting $L=1$. Figure \ref{TAdS3} shows the various phases marked from $I$ to $IV$ with corresponding representative potential plots. 

The sign of the slope of the potential at $\phi_{cl}=0$ depends only on $\b$. This is because the slope at $\phi_{cl}=0$ gets contribution from $e^{-n\b\D_v}$ and the term linear in $M_v$ from the zero temperature contribution. The first derivative of the potential at $\phi_{cl}=0$ equated to zero gives this value of $\b=\b_1$. Keeping only the first term $(n=1)$ in the finite temperature expansion the slope is zero when $(1-e^{-\b_1})^2=8 e^{-\b_1}$ which gives $\b_1 \sim 2.3$. The vertical line at $\b_1$ divides the two regions with positive or negative slope at $\phi_{cl}=0$. There is thus a second order transition across regions $I$ to $IV$.

The potential is smooth at $\phi_{cl}=0$ along the vertical line at $\b_1$ and moves over from a single to a double well as $m^2$ is decreased from positive to negative values. On either side of the dashed line the potential is non-differentiable at $\phi_{cl}=0$. 

In region $II$ a new set of minima exist at $\phi_{cl}=\pm\phi_0$. In region $III$, $V(\phi_0)<V(0)$. There is a first order transition across $II$ and $III$. 

We further note that due to the negative sign of the regularized volume for AdS$_3$ the $U(1)$ symmetry remains broken at high temperatures as in the case of theories with only scalars \cite{Kakkar:2022hub}. 

\subsubsection{AdS$_4$}\label{phases4v}

Expanding the zero temperature traces about $d=3$ with $\e=3-d$, from \cite{Kakkar:2022hub} we have for the scalars

\beqa\label{expads4s}
\frac{1}{{\cal V}^{}_{d+1}}\mbox{tr} \frac{1}{-\square+M_i^{2}}=\frac{(2+M_s^2)}{16\pi^2}\left[-\frac{2}{\e}-1+\gamma-\log(4\pi)+\psi^{(0)}\left(\n_i-\f{1}{2}\right)+\psi^{(0)}\left(\n_i+\f{3}{2}\right)\right]\non
\eeqa

and from (\ref{trzerotv4t}) we get the following  contribution from the transverse vector trace

\begin{eqnarray}\label{expads4v}
\f{1}{{\cal V}_{d+1}}\mbox{tr} \left[\frac{1}{\mathcal{O}^T+M_v^2-d+1}\right]=\frac{3}{16\pi^2}\left(\nu_v^2-\frac{9}{4}\right) \left[-\frac{2}{\epsilon}-1+\gamma-\log(4\pi)+2\psi^{(0)}\left(\frac{1}{2}+\nu_v\right)\right].
\end{eqnarray}

\vspace{0.5cm}
\noindent
{\bf Phases as a function of AdS radius}
\vspace{0.3cm}

We write down the potential in the $\overline{\mbox{MS}}$ scheme in the Landau gauge introducing the scale $\mu$ and putting back the AdS radius $L$ into the expression. 

\beqa\label{potEads4}
V_{eff}=\f{\l}{4}\phi_{cl}^4+\f{3}{32\pi^2 L^4}\int_0^{L^2M_v^2}(x^2-2)\left[2\psi^{(0)}(1/2+\sqrt{1/4+x^2})-1-\log(\m^2L^2)\right]dx^2.
\eeqa

In this scheme we have removed the finite $\g+\log(4\pi)$ along with the $1/\epsilon$ piece from (\ref{expads4v}). We have also ignored the contribution from the scalar loops as we will set $\l \sim e^4$ as discussed below. Before analyzing the potential numerically we now extract a few analytical results by looking at two regimes, large $L$ and small $\phi_{cl}$. To get the large $L$ potential we first expand the integrand in (\ref{potEads4}) for large values of $x$

\beqa
\psi^{(0)}(1/2+\sqrt{1/4+x^2})\sim \log x+\f{1}{6x^2}-\f{1}{30x^4}+{\cal O}(1/x^6)
\eeqa

\beqa\label{potLads4}
V_{eff}\sim\f{\l}{4}\phi_{cl}^4+\f{3M_v^4}{64\pi^2 }\left[\log(M_v^2/\m^2)-3/2\right]-\f{3M_v^2}{16 \pi^2 L^2}\left[\log(M_v^2/\m^2)-13/6\right] +{\cal O}(1/L^4).
\eeqa

The $L$ independent terms give the flat space result. The minimum of the flat space potential is at $\phi_0^2=\m^2/(2e^2)\exp(1-4\pi^2\l/3e^4)$. Setting $\mu^2=2e^2\phi_0^2$ gives $\l=3e^4/4\pi^2$. In the following analysis we shall use this value of $\l$ keeping  $\m^2$ fixed.

\begin{figure}[H] 
\begin{center} 
  \begin{minipage}{0.25\textwidth}%
   \begin{subfigure}[b]{0.98\linewidth}
    \centering
    \includegraphics[width=1\linewidth]{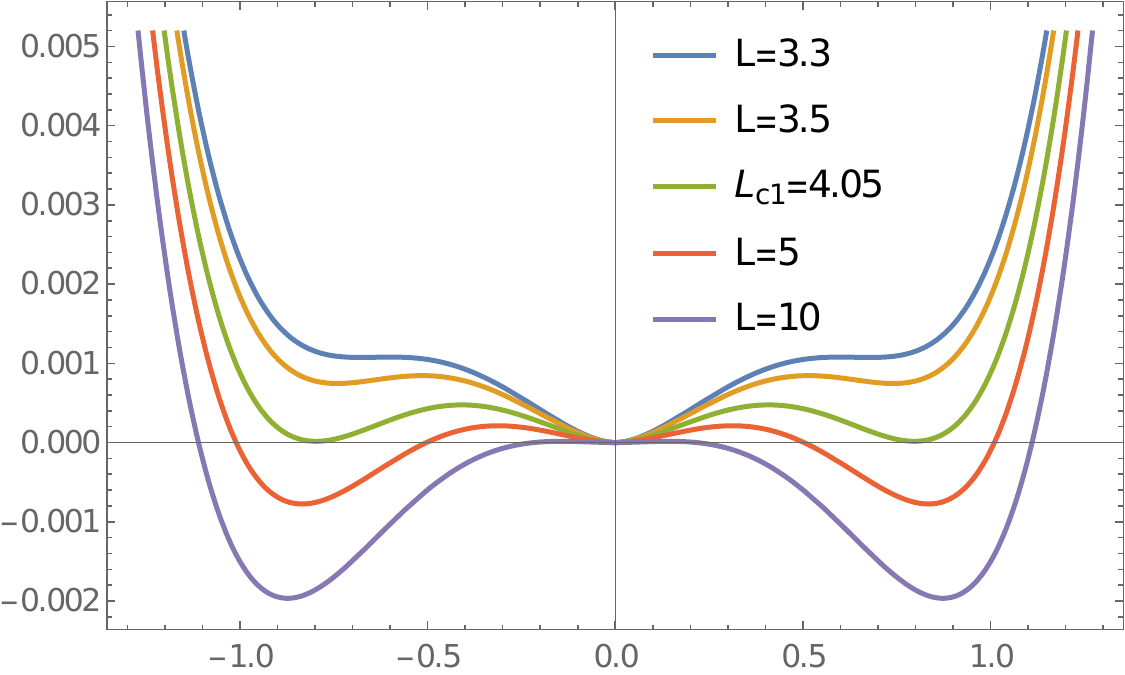} 
    \caption{} 
    \label{LAdS4} 
    \vspace{1ex}
  \end{subfigure}
   \begin{subfigure}[b]{0.98\linewidth}
    \centering
    \includegraphics[width=1\linewidth]{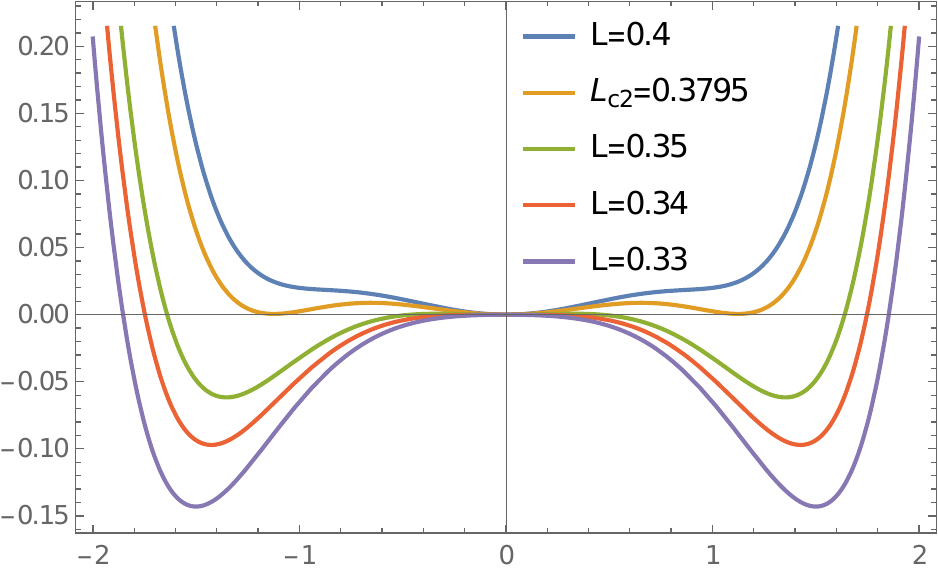} 
    \caption{} 
    \label{SAdS4} 
    \vspace{1ex}
  \end{subfigure}
\end{minipage}
\begin{minipage}{0.25\textwidth}%
   \begin{subfigure}[b]{0.98\linewidth}
    \centering
    \includegraphics[width=1\linewidth]{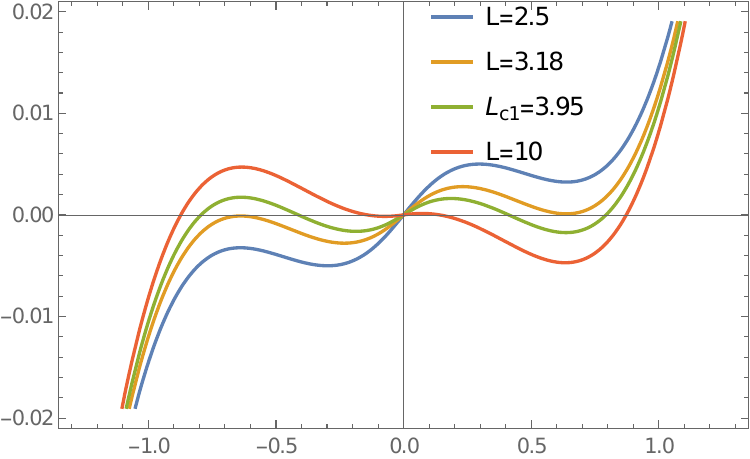}   
    \caption{} 
    \label{EAdS41} 
    \vspace{1ex}
  \end{subfigure}
  \begin{subfigure}[b]{0.98\linewidth}
    \centering
    \includegraphics[width=1\linewidth]{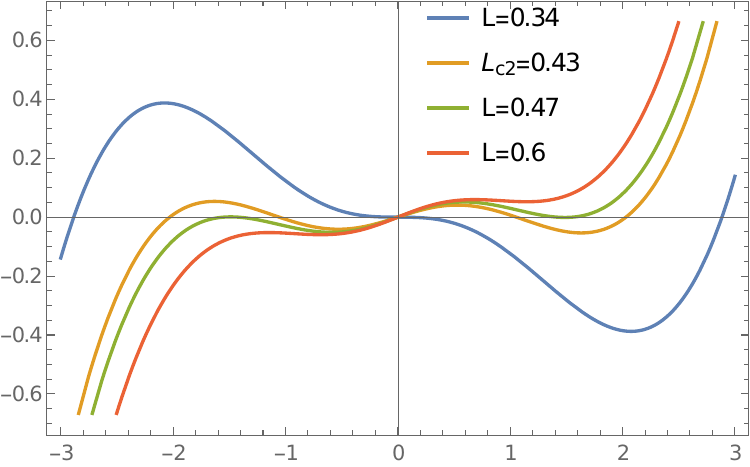}   
    \caption{} 
    \label{EAdS42} 
    \vspace{1ex}
  \end{subfigure}
  \end{minipage}
  \begin{minipage}{0.45\textwidth}  
   \begin{subfigure}[b]{0.90\linewidth}
    \centering
    \includegraphics[width=1\linewidth]{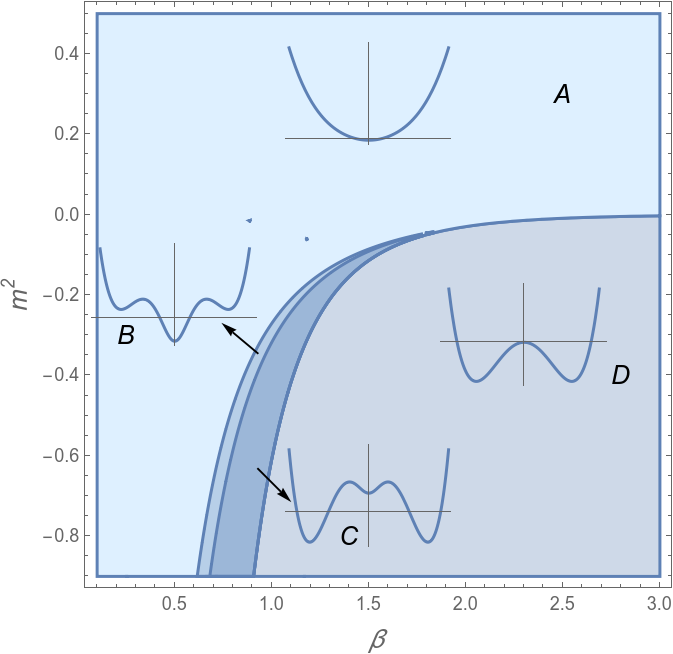}   
    \caption{} 
    \label{TAdS4} 
    \vspace{1ex}
  \end{subfigure}
  \end{minipage}
   \caption{(a) and (b) are plots of the effective potential $V_{eff}(\phi_{cl})$ for various values of AdS radius $L$ ($\m^2=1$). (a) for large $L$ limit (\ref{potLads4}) $(e=0.8)$ and (b) for small $\phi_{cl}$ (\ref{potSads4}) $(e=0.9)$ . (c) and (d) are derivatives to the potential (\ref{potEads4}) $(e=0.8)$. (e) is the phase plot on the $\b$-$m^2$ plane with $L=1$ and $e=0.6$.}
  \label{PAdS4v} 
  \end{center} 
\end{figure}

Since the quadratic term in  $\phi_{cl}$ in (\ref{potLads4}) has a positive coefficient, there is a minimum at $\phi_{cl}=0$. Two additional minima exist at $|\phi_{cl}|=\phi_0$ and as $L$ is increased there is a first order transition at a critical radius $L_{c1}$ where the minima at $\phi_{cl}=0$ and $\phi_{cl}=\phi_0$ exchange dominance. This is shown in figure (\ref{LAdS4}). $L_{c1}$ is obtained by solving simultaleously $V_{eff}(\phi_0,L_{c1})=V_{eff}(0,L_{c1})$ and $V^{\prime}_{eff}(\phi_0,L_{c1})=0$ with $V^{\prime\prime}_{eff}(\phi_0,L_{c1})> 0$. For $\l\sim e^4$, $L_{c1}$ is independent of $e$. The charge only plays the role of scaling the potential plots. The minimum of the potential $\phi_0$ at $L_{c1}$ which determines the vector mass is thus inversely proportional to $e$. As $L$ is increased to infinity the maxima merge with the minimum at $\phi_{cl}=0$ to give the flat space double well potential.

We next consider small values of $\phi_{cl}$. Using the following expansion for small values of $x$

\beqa
\psi^{(0)}(1/2+\sqrt{1/4+x^2})\sim -\g+\f{\pi^2x^2}{6}+(\psi^{(2)}(1)/2-\pi^2/6)x^4 +(\pi^2/3+\pi^4/90-\psi^{(2)}(1))x^6+{\cal O}(x^8)\non
\eeqa

we have the following effective potential,

\beqa\label{potSads4}
V_{eff}&\sim&\f{\l}{4}\phi_{cl}^4+\f{3}{32\pi^2 L^4}\left[(LM_v)^2\left(4\g+2+2\log(\m^2L^2)\right)+(LM_v)^4\left(-\g-1/2-\pi^2/3-\log(\m^2L^2)/2\right)\right.\non 
&+& \left.(LM_v)^6\left(\pi^2/3-2\psi^{(2)}(1)/3\right)\right]+{\cal O}((LM_v)^8).
\eeqa 

For $\m L<\m L_1=\exp(-\g-1/2)=0.34$ we have a maximum at $\phi_{cl}=0$ and a minimum otherwise. For $\m L>\exp(-(\g+\pi^2/3))=0.02$ the coefficient of the $\phi_{cl}^4$ term is negative. Thus the coefficient of the quartic term changes sign before the maximum at the origin turns into a minimum as $\m L$ is increased. We have thus included the $\phi_{cl}^6$ term. For $\m L<<1$ , we have a double well potential. As  $L$ is increased further, new minima appear followed by a first order transition at $L=L_{c2}$ as in the large $L$ case\footnote{In De Sitter space a similar analysis leads to a second order transition \cite{Shore:1979as}\cite{Allen:1983dg}.}.
$L_{c2}$ can be computed in the same way as mentioned earlier and is independent of $e$. The corresponding potentials are shown in figure (\ref{SAdS4}).  

Numerical analysis with the full potential (\ref{potEads4}) gives the sum of the two features discussed above. There are essentially two first order phase transitions as a function of $L$ at $L_{c1}=3.95$ and at $L_{c2}=0.43$. The approximate critical values of $L$ shown in figures \ref{LAdS4} and \ref{SAdS4} are close to these values. At intermediate values of $0.47<L<3.18$ the potential has a single minimum at $\phi_{cl}=0$. Figures \ref{EAdS41} and \ref{EAdS42} show the derivatives of the potential at various values of $L$. The roots of the first derivative are consistent with the extrema of the potential plots corresponding to the two approximations.   

\noindent
{\bf Phases at finite temperature}
\vspace{0.3cm}

We now study the phases as a function of $m^2$ and $\b$. For this it is easier to work by setting renormalization conditions as $\phi_{cl}=0$. In four dimensions two and four point functions in $\phi_{cl}$ are UV divergent. We use the renormalization scheme as used in \cite{Kakkar:2022hub}. 

The counterterms have the following form
\begin{equation}\label{counterv}
\phi_{cl}^2~\d m^2+\f{1}{4}\phi_{cl}^4~\d \l~.
\end{equation}

We use the following renormalization conditions at $\phi_{cl}=0$

\beqa\label{rcads4v}
\f{\pa^2 V^0_{eff}}{\pa \phi_{cl}^2}=2 m^2~~~~~~~~~~\f{\pa^4 V^0_{eff}}{\pa \phi_{cl}^4}=6\l.
\eeqa

and obtain the effective potential at zero temperature after also including the ghost trace, with $L=\xi=1$. The expression for the effective potential is given in appendix \ref{effads4v} equation (\ref{effp4v}).

Since the renormalization condition defining the finite coupling $\l$ is set at $\phi_{cl}=0$ we would not be able to capture the zero temperature phases as a function of $L$ as compared to the $\overline{\mbox{MS}}$ scheme used earlier where a finite renormalization scale $\m^2$ was introduced. In the absence of another scale in this scheme, the shape of the potential remains invariant with change in $L$. This scheme however has the benefit of having one less parameter. We thus use this zero temperature potential (\ref{effp4v}) for analysis of the full theory at finite temperature.  

Before analyzing the phases for this theory at finite temperature, let us note the flat space limit of the zero temperature effective potential in equation (\ref{effp4v}) which is obtained  by taking $L\rightarrow \infty$ limit. This translates into taking both $\phi_{cl}$ and $m^2$ to be large. Working in the Landau gauge with $\xi=0$, the fourth line in equation (\ref{effp4v}) does not contribute and $M_2^2=m^2+\frac{\l}{2}\phi_{cl}^2$, we get

\begin{eqnarray}
V_{\text{flat}}^0&=&m^2\phi_{cl}^2+\frac{\l}{4}\phi_{cl}^4+\frac{(m^2+\frac{3}{2}\l \phi_{cl})^2}{64 \pi^2} \log \left(1+\frac{m^2+\frac{3}{2}\l \phi_{cl}^2}{m^2}\right)\\ \nonumber
&+&\frac{(m^2+\frac{1}{2}\l \phi_{cl})^2}{64 \pi^2}\log \left(1+\frac{m^2+\frac{1}{2}\l \phi_{cl}^2}{m^2}\right)+\frac{3 (2 e^2 \phi_{cl}^2)^2}{64 \pi^2}\log \left(1+\frac{2 e^2 \phi_{cl}^2}{m^2}\right)\\ \nonumber
&-&\frac{\phi_{cl}^2}{2}\left(\frac{ e^2(-3 +16 \gamma)}{8 \pi^2}\right)-\frac{\phi_{cl}^4}{24}\left(-\frac{ e^4\left(3+18\gamma+10\pi^2\right)}{3\pi^2} +\frac{15 \l^2}{16 \pi^2}\right).
\end{eqnarray}

We note that with the renormalization conditions (\ref{rcads4v}) at $\phi_{cl}=0$, it is not possible to take the zero mass limit of the above effective potential because of the presence of the $\log(m^2)$ terms. To compare with the massless flat space theory we thus set the scalar mass to zero in the AdS effective action. Then the large $\phi_{cl}$ limit gives

\begin{eqnarray}
V_{\text{flat}}^0=\frac{\l}{4}\phi_{cl}^4+\frac{\phi_{cl}^4}{64 \pi^2}\left[\log \phi_{cl}^2 \left(\frac{5 \l^2}{2}+12 e^4 \right)+\frac{5\l^2}{9}\left(-\frac{21}{4}+9\gamma-\pi^2\right)+8 e^4\left(3\gamma+\pi^2-\frac{3}{4}\right) \right].\non
\end{eqnarray}

After the rescaling $\l \rightarrow (2/3)\l$ and $\phi_{cl}\rightarrow \phi_{cl}/\sqrt{2}$, we note that the leading order $\log$ term matches with Coleman-Weinberg potential in \cite{Coleman:1973jx,Weinberg:1973am}. The subleading, $\log$-independent $\phi^4_{cl}$ coefficient differs. This is because in our case it is the AdS effective potential which is renormalized at $\phi_{cl}=0$. This leads to difference in the finite terms as compared to a direct computation in a flat space.

The complete expression for the effective potential in AdS$_4$ at finite temperature is

\begin{eqnarray}\label{ads4ftv}
 V_{eff}= V^0_{eff}-\frac{3}{2\pi \b}\sum _{n=1}^{\infty} \frac{1}{ n (1-e^{ -n \b })^3}\left[ 3 e^{ - \b n \D_v}-e^{ - \b n \D_s}+e^{-\b n  \Delta_1}+e^{ -\b n \Delta_2}\right]
\end{eqnarray}
where
\begin{eqnarray}\label{deltaads4}
\D_v=\frac{3}{2} +\sqrt{\frac{1}{4}+ |M_v^2 |} ~~~~~
\D_s=\frac{3}{2}+\sqrt{\frac{9}{4}+M_v^2} ~~~~~
\D_i=\frac{3}{2}+\sqrt{\frac{9}{4}+M_i^2}.
\end{eqnarray}

The phases as a function of $\b$ and $m^2$ are shown in figure \ref{TAdS4}. The renormalization conditions (\ref{rcads4v}) imply that we have a minimum at $\phi_{cl}=0$ at zero temperature. For $L=1$ the phase transitions occur only when the other dimensionful parameter, $m^2$ is nonzero and negative. We have a line of first order transitions denoted by the boundary of regions $B$ and $C$. This line ends in a second order transition at zero temperatures. In the potential plot corresponding to region $C$, the maximum and the minimum at $\phi_{cl}=0$ merge to give the double well potential shown within region $D$. We note that for $L=1$ we can compare with the earlier analysis (in the $\overline{\mbox{MS}}$ scheme) shown in figures \ref{PAdS4v} (a)-(d) which was done with $\m^2=1$ at zero temperature. The massless theory lies in the region $A$ of figure \ref{TAdS4} which is consistent with the previous result.   

We end this section by giving some essential points on the numerics at finite temperature. The first derivative of the potential, $V^{\prime}(\phi_{cl})$ plotted against $\phi_{cl}$ has the shape as shown in figures \ref{EAdS41} and \ref{EAdS42}. New extrema appear when the minimum (on the righthand side) touches the $\phi_{cl}$ axis. Thus the phase boundary between regions $A$ and $B$ is obtained by solving simultaneously $V^{\prime}(\phi_0)=0$, $V^{\prime\prime}(\phi_0)=0$ along with the condition that $V^{\prime\prime\prime}(\phi_0)>0$, where $\phi_0\ne 0$. The boundary separating regions $B$ and $C$ is obtained by satisfying $V(\phi_0)=V(0)$ along with the above conditions. The complement of the union of regions $B$ and $C$ are the disjoint regions $A$ and $D$ which are distinguished by a minimum and maximum respectively at $\phi_{cl}=0$.  
 
\subsubsection{Gauge dependence}\label{gdep}

The effective potentials discussed in the previous sections are dependent on the gauge parameter $\xi$. However physical quantities derived from this potential should be gauge invariant. Studies of gauge dependence of effective action in flat space include \cite{Nielsen:1975fs}-\cite{Andreassen:2014eha}.  Using the Ward identity, in \cite{Nielsen:1975fs}, Nielsen derived a non-perturbative equation satisfied by the effective potential which is written as 
\beqa\label{neilsen}
\left[\xi\frac{\partial}{\partial \xi} +C(\phi_{cl},\xi)\frac{\partial}{\partial \phi_{cl}}\right] V_{eff}(\phi_{cl},\xi)=0.
\eeqa
This equation shows that the vacuum energy density or the value of the effective potential at an extremum is gauge invariant. $C(\phi_{cl},\xi)$ can be computed perturbatively and we leave this calculation for the theories studied in this paper as a future exercise\footnote{It needs to be shown that $C(\phi_{cl},\xi)$ at the extremum is finite.}. In the following, we show the gauge invariance of the minimum to order $\hbar$ for the theories in AdS. For this consider the following $\hbar$ expansions
\beqa
&&V_{eff}(\phi_{cl},\xi)=V_0+\hbar V_1(\phi_{cl},\xi)+\hbar^2 V_2(\phi_{cl},\xi)+\cdots\\
&&C(\phi_{cl},\xi)=C_0(\phi_{cl},\xi)+\hbar C_1(\phi_{cl},\xi)+\hbar ^2C_2(\phi_{cl},\xi)+\cdots~~.
\eeqa

Equating powers of $\hbar$ on both sides of (\ref{neilsen}) we have
\beqa\label{neilsen1}
\xi\frac{\partial }{\partial \xi}V_{1}(\phi_{cl},\xi) +C_1(\phi_{cl},\xi)\frac{\partial}{\partial \phi_{cl}} V_{0}(\phi_{cl})=0
\eeqa
where the fact that $V_0$ is independent of $\xi$ has been used.
Thus the one-loop potential is gauge-invariant whenever $\phi_{cl}$ at tree-level is the extremum value $v_0$. It is not difficult to check this for the theories discussed here. For the classical potential in equation (\ref{scalarqedL1})
\beqa
\phi_{cl}^2=v_{0}^2=-\frac{2 m^2}{\l}.
\eeqa

For this value of $\phi_{cl}$, the masses of the various fields defind in (\ref{massdefv}) become
\begin{eqnarray}
M^{2}_{1}=-2 m^2~~~~M^{2}_{2}=-\frac{4 \xi e^2 m^2}{\l} ~~~~~
M^{2}_{v}=-\frac{4 e^2 m^2}{\l}.
\end{eqnarray}
At this level the ghost and the scalar $\eta_2$ have the same mass. The longitudinal contribution is the same as that of a scalar of mass $\sqrt{\xi}M_v$. Thus from equation (\ref{potsqed}), the gauge parameter dependent quantities cancel and the minimum of the potential is gauge invariant.\footnote{Alternately, suppose $\phi_0$ is an extremum of $V_{eff}$. Then expanding in powers of $\hbar$,

\beqa
\phi_0=v_0+\hbar v_1 + \cdots ~~\mbox{and}~~
V_{eff}(\phi_0)=V_0(\phi_0)+\hbar V_1(\phi_0)+\cdots \sim  V_0(v_0)+\hbar V_1(v_0)\nonumber.
\eeqa
We can check the gauge invariance of the extremum by evaluating $V_1(v_0)$ explicitly as explained above.} This is valid for any general dimensions, $d$. For example in AdS$_3$, from (\ref{potzerot3v}) we get

\beqa
V^0_{eff}(v_0)=-\frac{m^4}{\l}+\frac{1}{6 \pi}\left[3 \left(\frac{-4 e^2 m^2}{\l} \right)^{1/2}-\left(\frac{-4 e^2 m^2}{\l} \right)^{3/2}\right]-\frac{(1-2m^2)^{3/2}}{12 \pi}.
\eeqa
Similar cancellation of the gauge parameter dependent quantities can be checked explicitly from the expression for AdS$_4$. In the finite temperature contributions given in equations (\ref{ads3ftv}) and (\ref{ads4ftv}), the gauge parameter dependent contributions to the effective potentials cancel at the one-loop level since $\Delta_s$ and $\Delta_2$ become equal but have contributions with opposite signs.

\subsection{$N$ scalars}\label{effpotNs}

In this section we derive effective potentials for theories with $N$ scalars. The expressions that we derive 
differ from those in the previous sections in the definition of the classical effective masses for scalars and factors of $N$. We begin by considering the following Lagrangian
\begin{equation}
\mathcal{L}=\frac{1}{4}F^{}_{\mu \nu}F^{\mu \nu}+\left(D_{\mu}\phi^{a}\right)\left(D^{\mu}\phi^{a}\right)^{*}+m^{2}\phi^{a}\phi^{a*}+\frac{\l}{4}\left[\phi^a \phi^{a*}\right]^2~~~~~\mbox{where}~~~a=1,\cdots,N~.
\end{equation}
Following the steps in section \ref{partv} and expanding about the classical vacuum expectation value as

\begin{equation}
\phi_{1}^a=\sqrt{2}\phi_{cl}^a+\eta_1^a ~~~~~\mbox{and}~~~~~ \phi_2^a=\eta_2^a
\end{equation}

we get the following Lagrangian that is quadratic in the fields after having introduced a gauge fixing term 

\begin{equation}
G=\frac{1}{\sqrt{\xi}}(\nabla_{\mu}A^{\mu}+\sqrt{2}\xi e \eta_2^a \phi^a_{cl})
\end{equation}

and the Fadeev-Popov Ghosts similar to the analysis in section \ref{partv} 

\begin{eqnarray}
\mathcal{L}&=&\frac{1}{4}F^{}_{\mu \nu}F^{\mu \nu}+\frac{1}{2\xi}\left(\nabla_{\mu}A^{\mu}\right)^2+\frac{1}{2}M_v^2A_{\mu}A^{\mu}  +\overline{c}\left[-\nabla^2+2\xi^2 e^2(\phi_{cl}^a)^2\right]c \\ \nonumber
&+&\frac{1}{2}\left(\pa_{\m}\eta_1^a\right)^2+\frac{1}{2}\left(\pa_{\m}\eta_2^a\right)^2+\frac{1}{2}m^2\left(\eta_1^a\right)^2+\frac{1}{2}m^2\left(\eta_2^a\right)^2 +\frac{\l}{4}\left[\left(\phi_{cl}^a\right)^2\left(\left(\eta_1^a\right)^2+\left(\eta_2^a\right)^2\right)+2\left(\phi_{cl}^a\eta_1^a\right)^2\right]
\\ \nonumber
&+&\xi e^2 (\eta_2^a\phi_{cl}^a)^2+m^2\left(\phi_{cl}^a\right)^2+\frac{\l}{4}\left(\phi_{cl}^a\right)^4. \nonumber
\end{eqnarray}

Setting $\phi_{cl}^{a}=(0,0,\cdots,\phi_{cl})$ gives a modified Klein-Gordon operator for $\eta_1^a$, $[-\square^{}_{E}+M^{2}_{1i}]$ with
\begin{equation}
M^{2}_{1i}=
\begin{cases}
\frac{\lambda}{2}\phi^{2}_{cl}+m^{2}, & \text{for}\ \eta^{1}_1\cdots\eta^{N-1}_1 \\
\frac{3\lambda}{2}\phi^{2}_{cl}+m^{2}, & \text{for}\ \eta_1^{N}
\end{cases} 
\end{equation}
while the effective masses for $\eta_2^N$ and the vector are the same as $M_2$ and $M_v$ respectively as defined in (\ref{massdefv}).
The other $\eta_2^a$ where $a=1,\cdots,N-1$ have the mass $M_{11}$. The effective potential thus becomes

\beqa
\label{potsqedN}
V^{}_{eff}(\phi^{}_{cl})&=&-\frac{1}{{\cal V}^{}_{d+1}}\left[\log Z_{\mbox{\tiny{gauge}}}^{(1)}+\log Z_{\mbox{\tiny{ghost}}}^{(1)}+\log Z_{\mbox{\tiny{scalars}}}^{(1)}\right]+V(\phi_{cl})\non
&=&\f{1}{{\cal V}_{d+1}}\left[\underbrace{\f{1}{2}\mbox{tr}\log[-\mathcal{O}^T+M_v^2-d+1]+\f{1}{2}\mbox{tr}\log[-\mathcal{O}^L+\xi M_v^2]}_{\tiny{\mbox{transverse + longitudinal vectors}}}\underbrace{-\mbox{tr}\log[-\square^{}_{E}+\xi M_v^2]}_{\tiny{\mbox{ghost}}}\right.\non
&+&\left.\underbrace{\frac{1}{2}\times 2 \left(N-1\right)\mbox{tr} \log[-\square^{}_{E}+M_{11}^2]}_{\eta^a_{1,2}, ~a=1,\cdots,N-1}+\underbrace{\frac{1}{2}\mbox{tr} \log[-\square^{}_{E}+M_{1N}^2]}_{\eta_1^N}\right.\non
&+&\left.\underbrace{\frac{1}{2}\mbox{tr} \log[-\square^{}_{E}+M_2^2]}_{\eta_2^N}\right]+V(\phi^{}_{cl}).
\eeqa

A similar method as discussed in section \ref{gdep} can be used to show that the value of the potential at an extremum is gauge invariant.

\subsubsection{AdS$_3$}

The effective potential for the finite $N$ case, including the expression
for the transverse vector trace from (\ref{trzerotv4t}) and the scalar trace from \cite{Kakkar:2022hub}, can be written as

\begin{eqnarray}\label{potzt3n}
V_{eff}^0&=&m^2\phi_{cl}^2+\frac{\l}{4}\phi_{cl}^4+\frac{3(M_v^2)^{1/2}-(M_v^2)^{3/2}}{6 \pi}+\frac{\left(1+\xi M_v^2\right)^{3/2}}{12\pi}\\ \nonumber
&-&\frac{1}{12\pi}\left[2(N-1)\left(1+M_{11}^2\right)^{3/2}+\left(1+M_{1N}^2\right)^{3/2}\right]-\frac{\left(1+M_2^2\right)^{3/2}}{12\pi}.
\end{eqnarray}

For $N = 1$ the expression for the effective potential reduces to that of the single scalar case as described in equation (\ref{potzerot3v}). 

\begin{figure}[H] 
\begin{center} 
  \begin{minipage}{0.32\textwidth}%
   \begin{subfigure}[b]{0.9\linewidth}
    \centering
    \includegraphics[width=1\linewidth]{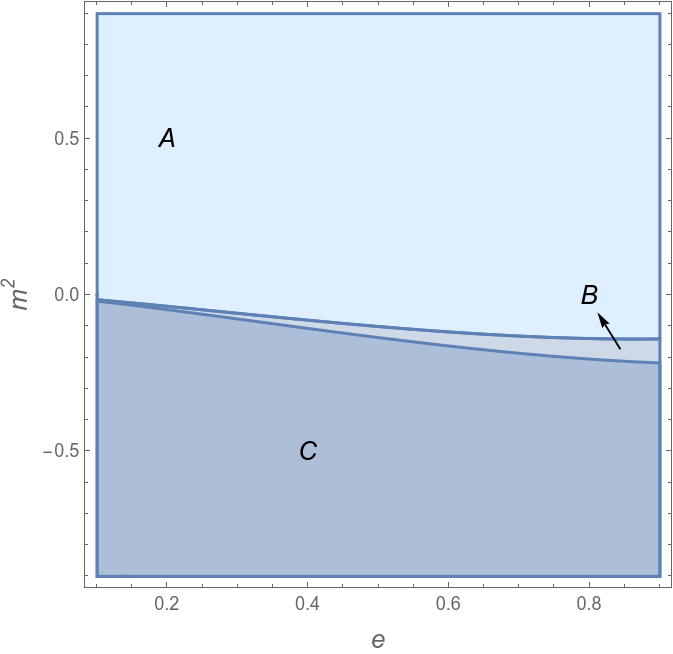} 
    \caption{} 
    \label{EAdSN1} 
    \vspace{1ex}
  \end{subfigure}
  \end{minipage}
 \begin{minipage}{0.32\textwidth}%
   \begin{subfigure}[b]{0.9\linewidth}
    \centering
    \includegraphics[width=1\linewidth]{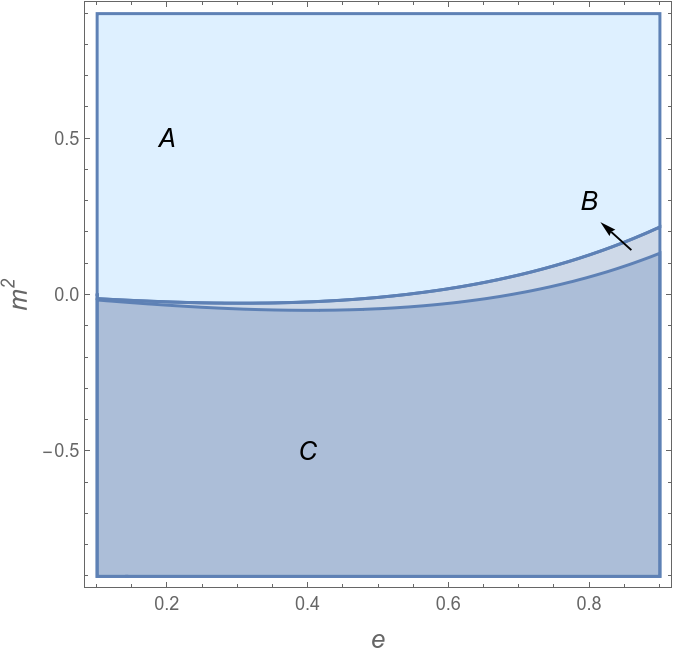} 
    \caption{} 
    \label{EAdSN10} 
    \vspace{1ex}
  \end{subfigure}
\end{minipage}
\begin{minipage}{0.32\textwidth}%
   \begin{subfigure}[b]{0.9\linewidth}
    \centering
    \includegraphics[width=1\linewidth]{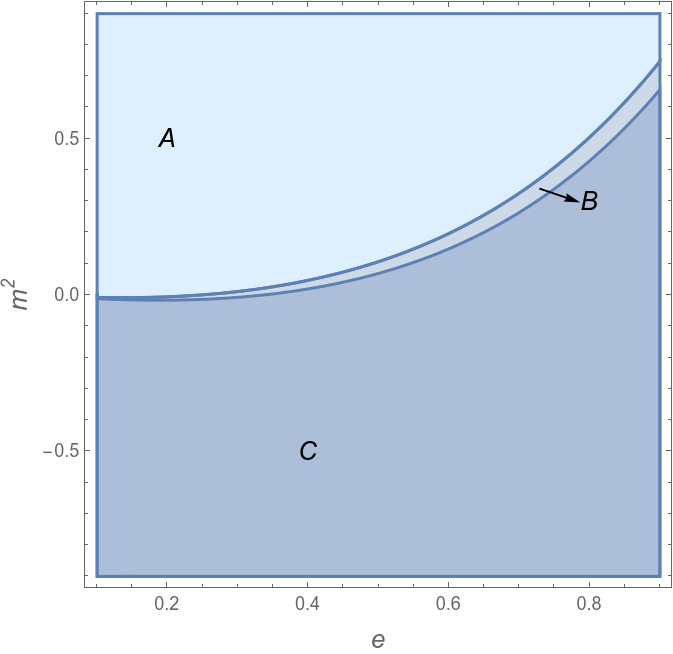} 
    \caption{} 
    \label{EAdS3N20} 
    \vspace{1ex}
  \end{subfigure}
\end{minipage}
  \caption{Phases as a function of $e$ and $m^2$ at zero temperature with $L=1$ for (a) $N=1$ (b) $N=10$ (c) $N=20$.}
  \label{EAdS3N} 
  \end{center} 
\end{figure}

In figure \ref{EAdS3N} the phases at zero temperature are plotted as a function of $m^2$ and $e$ setting the AdS radius $L=1$ for various values of the number of scalars, $N$. The figure shows three regions, $A$, $B$ and $C$ which correspond to potentials as in figure \ref{EAdS3}. Thus in the numerical plots the first order transition exists for all the values of $N$ studied and is expected to be so for all finite values of $N$. 

At finite temperature, collecting the expressions from (\ref{transtt3}), (\ref{longtt3}) the potential is

\beqa
V_{eff}=V_{eff}^0+\f{2}{\pi\b}\sum_{n=1}^{\infty}\f{1}{n(1-e^{-n\b})^2}\left[2e^{-n\b\D_v}-e^{-n\b\D_s}+2(N-1)e^{-n\b\D_{11}}+e^{-n\b\D_{1N}}+e^{-n\b\D_2}\right]\non
\eeqa
where $\D_v$, $\D_s$, $\D_2$ are definied in (\ref{deltaads3}) and $\D_{1i}=1+\sqrt{1+M_{1i}^2}$.

The $\b-m^2$ phase plots for finite $N$ are qualitatively similar to that of a single scalar in figure \ref{TAdS3}.  

\subsubsection{AdS$_4$}

Including counterterms as in (\ref{counterv}) and imposing the renormalization conditions (\ref{rcads4v}) we can write down the expression for the zero temperature effective potential from (\ref{potsqedN}). The potential is given in appendix \ref{effads4v} equation (\ref{sqedads4N}). The renormalization condition (\ref{rcads4v}) implies that the zero temperature potential has a maximum for $m^2<0$. Since $V^0(0)=0$ we have a symmetry breaking only when the tree-level scalar mass-squared, $m^2<0$ for all finite values of $N$ and $0 < e \le 0.9$. 

It is easy to check that for $N = 1$ the expression for the effective potential again reduces to that of the single scalar case as in (\ref{effp4v}). The complete expression for the effective potential in AdS$_4$ at finite temperature is

\begin{eqnarray}
 V_{eff}=V^0_{eff}-\frac{3}{2\pi \b}\sum _{n=1}^{\infty} \frac{1}{ n (1-e^{ -n \b })^3}\left[ 3 e^{ - \b n \D_v}-e^{ - \b n \D_s}+2(N-1)e^{-\b n  \Delta_{11}}+ e^{-\b n  \Delta_{1N}}+e^{ -\b n \Delta_2}\right]\non
\end{eqnarray}
where $\D_v$, $\D_s$, $\D_2$ are definied in (\ref{deltaads4}) and $\D_{1i}=3/2+\sqrt{9/4+M_{1i}^2}$.

In the large $N$ limit, to the leading order, the vectors play no role in the determination of the phases. The phases are those of the $O(2N)$ scalar model studied earlier in \cite{Kakkar:2022hub}.

\section{Discussion}\label{summaryv}

We end this paper with a discussion on some future directions. In this work we studied phases of scalar QED theories as a function of AdS radius for zero temperature and further as function of scalar mass and temperature including only the Maxwell term. As extension of this work we would like to explore theories with Chern-Simons term as well, including theories with non-Abelian gauge symmetry. 

A large portion of this paper along with \cite{Kakkar:2022hub}\cite{Kakkar:2023gzu} is devoted to the derivation of one loop partition function for spins zero, half and one by introducing an alternate method. It was shown for the case of scalars in \cite{Kakkar:2022hub} that the computation technique implemented for thermal partition function is equivalent to the method of images. As emphasized before, many of the results are already known in the literature. It will be interesting to explore the connections between these methods. The connection between the methods using heat kernel \cite{Giombi:2008vd}-\cite{Gupta:2012he} and normal modes \cite{Denef:2009kn} is explored in \cite{Martin:2019flv}. A further natural extension of our alternate method is to the case of higher spins. 
 
Another aspect which we would like to focus on is the computation of correlation functions for QFTs in thermal AdS with spherical spatial boundary. While our works involved computation of two point functions at coincident AdS spatial points, two and higher point functions can be computed in general. These have been computed on Euclidean AdS for various theories in \cite{Carmi:2018qzm}-\cite{Ankur:2023lum} and \cite{Bertan:2018khc}, \cite{Bertan:2018afl}. For AdS with flat boundary, thermal correlators were studied in \cite{Alday:2020eua}. These functions at finite temperature contain a large amount of information regarding the dynamical aspects of phase transition specially at the critical points. Such critical theories in Euclidean AdS have been studied recently in \cite{Carmi:2018qzm}-\cite{Giombi:2021cnr}. 

We hope to address these above mentioned areas in our future work.

\vspace{2.5mm}
\noindent
{\bf Acknowledgements :}
\\ We thank an anonymous referee of the paper for valuable comments and suggestions.
Astha Kakkar acknowledges the support of Department of Science and Technology
(DST), Ministry of Science and Technology, Government of India, for the DST
INSPIRE Fellowship with the INSPIRE Fellowship Registration Number: IF180721.
The work of S.S. is partially supported by the Personal Research Grant 2023-2024,
Vidyasagar University. S.S. also thanks the University Grants Commission (UGC),
and DST, New Delhi, India, for providing special assistance and infrastructural support
to the Department of Physics, Vidyasagar University, through the SAP and FIST
program respectively.

\appendix

\section{Some details of zero temperature computation}\label{detailsv}

\subsection{Normalizations}\label{normav}

To obtain the normalization we consider 

\beqa
\int d^{d+1}x~\sqrt{g}g^{\m\n}A^*_{\m}A_{\n}=\int d^{d+1}x~\sqrt{g}\tilde{A}^*_{\m}\tilde{A}_{\m}.
\eeqa

The above integral for the transverse vector $I=1$ is written as

\beqa\label{normtI1}
&&\int d^{d+1}x\sqrt{g}~[\psi^{T}_{\vec{k}^{\prime},\l^{\prime}}(\vec{x}, y)]^{\dagger}_{(1)}[\psi^T_{\vec{k},\l}(\vec{x}, y)]_{(1)}
\\
&=&
\f{1}{|\a_T|^2}\int \f{dy}{y^{d+1}}(ky)^d\left[(ky)^2\left\{K_{i\l}(ky)K_{-i\l^{\prime}}(ky)+K_{i\l+1}(ky)K_{-i\l^{\prime}+1}(ky)\right\}\right.\\
&-&\left.(ky)\left\{\a_T K_{i\l}(ky)K_{-i\l^{\prime}+1}(ky)+\a_T^*K_{-i\l^{\prime}}(ky)K_{i\l+1}(ky)\right\}
+ |\a_T|^2K_{i\l}(ky)K_{-i\l^{\prime}}(ky)  \right](2\pi)^d\d^d(\vec{k}-\vec{k^{\prime}})\non
&=& \f{1}{|\a_T|^2}\lim_{a \to 1} \int dy \left[y^a\left\{K_{i\l}(y)K_{-i\l^{\prime}}(y)+K_{i\l+1}(y)K_{-i\l^{\prime}+1}(y)\right\}\right.\\
&-&\left.\ y^{a-1}\left\{\a_T K_{i\l}(y)K_{-i\l^{\prime}+1}(y)+\a_T^*K_{-i\l^{\prime}}(y)K_{i\l+1}(y)\right\}
+ |\a_T|^2 y^{a-2} K_{i\l}(y)K_{-i\l^{\prime}}(y)  \right] k^d (2\pi)^d\d^d(\vec{k}-\vec{k^{\prime}}).\nonumber
\eeqa

$\a_T=(i\l-d/2+1)$. Evaluation of the $y$ integral on the r.h.s. of the above equation gives

\beqa
&&\f{1}{|\a_T|^2}\lim_{a \to 1}\frac{2^{a-6}}{\Gamma (a-1)} \left(2 a^2+2 a (d-4)+(d-6) d+2 \left(\l^2+\l^{\prime 2}+5\right)\right)\Gamma \left(\f{a-1}{2} +\f{i}{2}(\l-\l^{\prime})\right)\non &\times& \Gamma \left(\f{a-1}{2} -\f{i}{2}(\l-\l^{\prime})\right)\Gamma \left(\f{a-1}{2} +\f{i}{2}(\l+\l^{\prime})\right)\Gamma \left(\f{a-1}{2} -\f{i}{2}(\l+\l^{\prime})\right).
\eeqa

For $\l\ne \l^{\prime}$ the above expression is zero due to the presence of $\Gamma (a-1)$ in the denominator. Further if $\l \rightarrow \l^{\prime}$ also the expression reduces to,

\beqa\label{limitat}
\f{2^{-5}}{|\a_T|^2}|\Gamma(i\l)|^2 (4\l^2+(d-2)^2)\lim_{a \to 1}\f{4(a-1)}{(a-1)^2+(\l-\l^{\prime})^2}=\d(\l-\l^{\prime})/\m(\l).
\eeqa

Along with the momentum delta function this gives the r.h.s of (\ref{normt}). $\mu(\l)$ is defined in (\ref{mudef}).

In the same way for each of $I=2,\cdots,d$ we have 

\beqa
&&\int d^{d+1}x\sqrt{g}~[\psi^{T}_{\vec{k}^{\prime},\l^{\prime}}(\vec{x}, y)]^{\dagger}_{(I)}[\psi^T_{\vec{k},\l}(\vec{x}, y)]_{(I)}
\\
&=&
\int \f{dy}{y^{d+1}}(ky)^d\left[K_{i\l}(ky)K_{-i\l^{\prime}}(ky))\right](2\pi)^d\d^d(\vec{k}-\vec{k^{\prime}}).
\eeqa

The computation of normalization is same as the scalar field normalization \cite{Kakkar:2022hub} and gives the same result as (\ref{normt}).

The normalization for the longitudinal vector proceeds exactly as the transverse $I=1$ case (\ref{normtI1}). For the vector (\ref{longv}), the $y$ integral in this case gives

\beqa
&&\f{1}{|\a_L|^2}\lim_{a \to 1}\frac{2^{a-6}}{\Gamma (a-1)}\left(2 a^2-2 a (d+2)+d^2+2 \left(d+\l^2+\l^{\prime 2}+1\right)\right)\Gamma \left(\f{a-1}{2} +\f{i}{2}(\l-\l^{\prime})\right)\non &\times& \Gamma \left(\f{a-1}{2} -\f{i}{2}(\l-\l^{\prime})\right)\Gamma \left(\f{a-1}{2} +\f{i}{2}(\l+\l^{\prime})\right)\Gamma \left(\f{a-1}{2} -\f{i}{2}(\l+\l^{\prime})\right).
\eeqa

$\a_L=(i\l+d/2)$. This expression including momentum delta function gives the normalization (\ref{norml}) following (\ref{limitat}).

\subsection{Integral over $\l$}\label{lintegralv} 

From equation (\ref{trzerotv3t}) we have 

\begin{equation}
\tr\left[\frac{1}{-{\cal O}^T+M_v^2-d+1}\right]=\frac{{\cal V}_{d+1}(d)}{(4\pi)^{(d+1)/2}\G\left(\frac{d+1}{2}\right)}\int_{-\infty}^{\infty}\f{d\l }{\l^2+\n^2}\frac{|\G\left(\frac{d-2}{2}+i\l\right)|^2}{|\G\left(i\l\right)|^2}\left[\l^2+\left(d/2\right)^2\right]. 
\end{equation}

We perform the $\lambda$ integral by closing the contour in the upper half plane. In the UHP there are two sets of poles.

\begin{enumerate}
\item $\l^2+\nu^2=0 \rightarrow \l=+ i\nu$

\begin{equation}\label{res1}
[2\pi i \mbox{Res}f(\l)]_1 =\frac{\pi}{\nu}\frac{1}{\Gamma\left(\pm \nu\right)}\left[\frac{d}{2}\pm \nu\right]\Gamma\left(\frac{d}{2}-1\pm\nu\right)
\end{equation}

\item $\frac{d-2}{2}+ i\l=-n \rightarrow \l=i\left(n+\frac{d-2}{2}\right)$

\begin{eqnarray}\label{res2}
[2\pi i \mbox{Res}f(\l)]_2 &=&\left(2\pi i\right)\sum_{n=0}^{\infty}\left(\frac{1}{\nu^2-\left[\left(n+\frac{d-2}{2}\right)\right]^2}\right)\left(\left(\frac{d}{2}\right)^2-\left[\left(n+\frac{d-2}{2}\right)\right]^2\right) \non
&\times & \Gamma\left(d-2+n\right)\frac{(-1)^n}{n!}i\left(n+\frac{d-2}{2}\right)\frac{\sin\left(\pi\left[n+\frac{d-2}{2}\right]\right)}{\pi} \\ \non
&=&-\frac{\left(d^2-4\nu^2\right)\pi^2\cos(\nu\pi)\tan\left(\frac{\pi d}{2}\right)}{2\left[\cos(\pi d)-\cos(2\nu\pi)\right]\Gamma\left(2-\frac{d}{2}\pm \nu\right)}.
\end{eqnarray}
\end{enumerate}

Adding equations \ref{res1} and \ref{res2} gives
\begin{equation}
\left[\nu^2-\left(\frac{d}{2}\right)^2\right]\frac{\Gamma\left(\frac{d}{2}-1\pm \nu\right)\Gamma\left(\frac{1}{2}\pm\frac{d}{2}\right)}{\Gamma\left(\nu-\frac{d}{2}\right)\Gamma\left(1-\nu+\frac{d}{2}\right)}.
\end{equation}
Putting in all the other factors gives the final result
\begin{equation}\label{trzerotv}
\mbox{tr}\left[ \frac{1}{-\mathcal{O}^T+M_v^2-d+1}\right]=\frac{\mathcal{V}^{}_{d+1} d}{\left(4\pi\right)^{(d+1)/2}}\frac{\left[\nu^2-\left(\frac{d}{2}\right)^2\right]\Gamma\left(\frac{1}{2}-\frac{d}{2}\right)\Gamma\left(\frac{d-2}{2}\pm \nu\right)}{\Gamma\left(\nu-\frac{d}{2}\right)\Gamma\left(1-\nu+\frac{d}{2}\right)}.
\end{equation}

\section{Some details of finite temperature computations}\label{acomptv}

\subsection{Normalizations}

\noindent
{\bf The other transverse vector for} AdS$_3$

The transverse vector (\ref{trans23}) in the complex field notation (\ref{complnv}) is

\beqa
[\psi^{T}_{\vec{k},\l}(\vec{x},y)]_{(2)}=\left(\begin{array}{c}\tilde{A}_y^T(y,z)\\ \mathsf{A}^T(y,z)\end{array}\right)_{(2)}=\left(\begin{array}{c}0\\ -i\mathsf{k}/k\end{array}\right)\phi_{\vec{k},\l}(\vec{x},y).
\eeqa

The corresponding function invariant under thermal $(\g^{n})$ identification is

\beqa\label{trans2t}
[\Psi^{T}_{\vec{k},\l}(\vec{x},y)]_{(2)}=\f{1}{{\cal N}}\sum_{n=-\infty}^{\infty} \g^n[\psi^{T}_{\vec{k},\l}(\vec{x},y)]_{(2)}= \f{1}{{\cal N}}\sum_{n=-\infty}^{\infty} \left(\begin{array}{c}0\\ e^{i n\th}(-i\mathsf{k}/k)\end{array}\right)\phi_{\vec{k},\l}(\vec{x},y)
\eeqa

\beqa
&&\int d^{3}x~\sqrt{g}~[\Psi^{T}_{\vec{k^{\prime}},\l^{\prime}}(\vec{x},y)]^{\dagger}_{(2)}[\Psi^T_{\vec{k},\l}(\vec{x},y)]_{(2)}\\
&=&\f{1}{{\cal N}^2}\sum_{n,n^{\prime}}\int d^{3}x~\sqrt{g}\left[\tilde{A}^{T*}_y(\g^{n^{\prime}}(y,z))\tilde{A}^{T}_y(\g^n(y,z))+e^{i(n-n^{\prime})\th}\bar{\mathsf{A}}^T(\g^{n^{\prime}}(y,z))\mathsf{A}^T(\g^n(y,z))\right]\\
&=&\f{1}{{\cal N}^2}\sum_{n,n^{\prime}}\int\f{dy}{y}\left[
e^{i(n-n^{\prime})\th}\f{\bar{\mathsf{k}}^{\prime}\mathsf{k}}{k^{\prime}k}\left\{(ke^{-n\b})(k^{\prime}e^{-n^{\prime}\b})K_{-i\l^{\prime}}(k^{\prime}ye^{-n^{\prime}\b})K_{i\l}(kye^{-n\b})\right\}\right]
(2\pi)^2\d^2(\g^n\vec{k}-\g^{n^{\prime}}\vec{k}^{\prime})\non
&=& \f{1}{{\cal N}}\sum_{n}(ke^{-n\b})^2(2\pi)^2\d^2(\g^{n} \vec{k}-\vec{k}^{\prime})\d(\l-\l^{\prime})/\m(\l).\label{normI2v}
\eeqa

\noindent
{\bf The longitudinal vector for} AdS$_3$

We first rewrite the longitudinal vector (\ref{longv}) in terms of complex fields $\mathsf{A}^L$ as has been done for the transverse vectors.

\beqa
&&\int d^{3}x~\sqrt{g}~[\Psi^{L}_{\vec{k^{\prime}},\l^{\prime}}(\vec{x},y)]^{\dagger}[\Psi^L_{\vec{k},\l}(\vec{x},y)]\non
&=&\f{1}{{\cal N}^2}\sum_{n,n^{\prime}}\int d^{3}x~\sqrt{g}\left[\tilde{A}^{L*}_y(\g^{n^{\prime}}(y,z))\tilde{A}^{L}_y(\g^n(y,z))+e^{i(n-n^{\prime})\th}\bar{\mathsf{A}}^L(\g^{n^{\prime}}(y,z))\mathsf{A}^L(\g^n(y,z))\right]\\
&=&\f{1}{{\cal N}^2}\sum_{n,n^{\prime}}\int\f{dy}{y}(ke^{-n\b})(k^{\prime}e^{-n^{\prime}\b})\left[\f{1}{|\a_L(\l^{\prime})|}\left\{\a_L^*(\l^{\prime})K_{-i\l^{\prime}}(k^{\prime}ye^{-n^{\prime}\b})-(k^{\prime}ye^{-n^{\prime}\b})K_{-i\l^{\prime}+1}(k^{\prime}ye^{-n^{\prime}\b})\right\}\right.\label{normlv1}\non
&\times&\left. \f{1}{|\a_L(\l)|}\left\{\a_L(\l)K_{i\l}(kye^{-n\b})-(kye^{-n\b})K_{i\l+1}(kye^{-n\b})\right\}\right.\non
&+&\left.e^{i(n-n^{\prime})\th}\f{\bar{\mathsf{k}}^{\prime}\mathsf{k}}{k^{\prime}k}\left\{(ke^{-n\b}y)(k^{\prime}e^{-n^{\prime}\b}y)K_{-i\l^{\prime}}(k^{\prime}e^{-n^{\prime}\b})K_{i\l}(ke^{-n\b}y)\right\}\right]
\times(2\pi)^2\d^2(\g^n\vec{k}-\g^{n^{\prime}}\vec{k}^{\prime})\label{normlv2}\\
&=& \f{1}{{\cal N}}\sum_{n}(ke^{-n\b})^2(2\pi)^2\d^2(\g^{n} \vec{k}-\vec{k}^{\prime})\d(\l-\l^{\prime})/\m(\l)\label{normlv3}\\&=&(2\pi)^2k^2\d^2(\vec{k}-\vec{k}^{\prime})\d(\l-\l^{\prime})/\m(\l)\label{normlv4}.
\eeqa

The simplifications that are involved in going from (\ref{normlv1}) to (\ref{normlv4}) are same as those described below (\ref{normI1v4}).

\subsection{Traces}

\noindent
{\bf The other transverse vector for} AdS$_3$

\beqa\label{tracevt2}
&&\tr\left[\frac{1}{-{\cal O}^T+M_v^2-d+1}\right]_{(2)}=
\int d^{3}x\sqrt{g}\int \f{d^2k}{(2\pi)^2}\f{1}{k^2}\int_{0}^{\infty}\f{d\l ~\m(\l)}{\l^2+\n^2_v}[\Psi^{T}_{\vec{k},\l}(\vec{x},y)]^{\dagger}_{(2)}[\Psi^T_{\vec{k},\l}(\vec{x},y)]_{(2)}\non
&=& \f{1}{{\cal N}^2}\sum_{n,n^{\prime}}\int \f{dy}{y}\int\f{d^2k}{(2\pi)^2}\f{1}{k^2}\int\f{d\l ~\m(\l)}{\l^2+\n^2_v}e^{i(n-n^{\prime})\th}\bar{\mathsf{A}}^T(\g^{n^{\prime}}(y,z))\mathsf{A}^T(\g^n(y,z))\\
&=&
\f{1}{{\cal N}^2}\sum_{n,n^{\prime}}\int \f{dy}{y}\int\f{d^2k}{(2\pi)^2}\int\f{d\l ~\m(\l)}{\l^2+\n^2_v}e^{-(n+n^{\prime})\b}
\left[e^{(n-n^{\prime})\th} K_{i\l}(kye^{-n\b})K_{-i\l}(kye^{-n^{\prime}\b})\right]\non
  &\times& (2\pi)^2\d^2(\g^{n}\vec{k}-\g^{n^{\prime}}\vec{k}).
\eeqa

Beyond this, following the same steps as for the first transverse vector described below (\ref{tracevt}), we get the same result as (\ref{tracevtt1}).

\noindent
{\bf The longitudinal vector for} AdS$_3$

\beqa\label{longvt}
&&\tr\left[\frac{1}{-{\cal O}^L+M_v^2}\right]=
\int d^{3}x\sqrt{g}\int \f{d^2k}{(2\pi)^2}\f{1}{k^2}\int_{0}^{\infty}\f{d\l ~\m(\l)}{\l^2+\n^2_v}[\Psi^{L}_{\vec{k},\l}(\vec{x},y)]^{\dagger}[\Psi^L_{\vec{k},\l}(\vec{x},y)]\non
&=&
\f{1}{{\cal N}^2}\sum_{n,n^{\prime}}\int \f{dy}{y}\int\f{d^2k}{(2\pi)^2}\int\f{d\l ~\m(\l)}{\l^2+\n^2_v}\f{1}{|\a_L|^2}e^{-(n+n^{\prime})\b}
\left[|\a_L|^2 K_{i\l}(kye^{-n\b})K_{-i\l}(kye^{-n^{\prime}\b})\right.\non
&+&\left.(kye^{-n\b})(kye^{-n^{\prime}\b})\left\{e^{(n-n^{\prime})\th}K_{i\l}(kye^{-n\b)}K_{-i\l}(kye^{-n^{\prime}\b})
+K_{i\l+1}(kye^{-n\b})K_{-i\l+1}(kye^{-n^{\prime}\b})\right\}\right.\non
&-&\left.(ky)\left\{\a_L e^{-n^{\prime}\b}K_{i\l}(kye^{-n\b})K_{-i\l+1}(kye^{-n^{\prime}\b})+\a_L^*e^{-n\b}K_{-i\l}(kye^{-n^{\prime}\b})K_{i\l+1}(kye^{-n\b})\right\}
  \right]\non
  &\times& (2\pi)^2\d^2(\g^{n}\vec{k}-\g^{n^{\prime}}\vec{k}).
  \eeqa

We next use the identity (\ref{di2v}) and follow the same steps as for the transverse vector. The delta function imposes the limits defined in (\ref{limitsv}). The difference here is that the term that survives is independent of $\th$ which is the first term in the second line of the above equation. This gives 

\beqa
\tr\left[\frac{1}{-{\cal O}^L+M_v^2}\right]=\sum_{n=1}\f{e^{-n\b(1+\n_v)}}{\n_v|1-e^{2\pi i n\t}|^2}.
\eeqa

\noindent
{\bf Computation for} AdS$_4$

 The contribution from $[\Psi^{T}_{\vec{k},\l}(\vec{x},y)]_{(2)}$ is similar to (\ref{trevendv1}) with $i=1$ only, $d=3$ and 

\beqa
f_{(1)}(\th_1,\th_2) \rightarrow f_{(1)}(\th_1)=\left(|\mathsf{k}_1|^2\cos (n\th_1)+k_3^2\right)/k^2.
\eeqa 

The other transverse vectors (\ref{trans24}) in the complex notation with $\mathsf{k}_1=k_1+ik_2$ are

\vspace{-0.5cm}
\beqa\label{ads4t2}
[\psi^{T}_{\vec{k},\l}(\vec{x},y)]^t_{(2)}=\left(\begin{array}{ccc}0,& -i\mathsf{k}_1/|\mathsf{k}_1|,&0 \end{array}\right)\phi_{k,\l}(\vec{x},y)
\eeqa
\vspace{-1cm}
\beqa\label{ads4t3}
[\psi^{T}_{\vec{k},\l}(\vec{x},y)]^t_{(3)}=\left(\begin{array}{ccc}0,&k_3\mathsf{k}_1|/(|\mathsf{k}_1|k),&-|\mathsf{k}_1|/k\end{array}\right)\phi_{k,\l}(\vec{x},y).
\eeqa

From these we get a similar expression as (\ref{trevendv2}) with 

\beqa
f_{(4)}(\th_1,\th_2) \rightarrow f_{(3)}(\th_1)= \left(k_3^2\cos (n\th_1)+|\mathsf{k}_1|^2\right)/k^2.
\eeqa

Adding these and noting that $k^2=|\mathsf{k}_1|^2+k_3^2$ we get

\beqa
\log Z_{\t}^T&=& \sum_{n=1}^{\infty}\left[2\cos(n\th_1)+1\right]\f{e^{-n\b(\n_v+1/2)}}{n(1-e^{-n\b})}
\f{e^{-n\b}}{|1-e^{2\pi in\t_1}|^2}.
\eeqa

\section{Effective potentials for AdS$_4$}\label{effads4v}

\noindent

{\bf For single scalar}

\begin{eqnarray}\label{effp4v}
 V^0_{eff}&=& m \phi_{cl}^2+\frac{\l \phi_{cl}^4}{4}\\ \nonumber
&+&\frac{1}{32 \pi^2}\int^{M_1^2}_{m^2}\left(M_1^2+2\right) \left(\psi^{(0)} \left(\sqrt{M_1^2+\frac{9}{4}}+\frac{3}{2}\right)+\psi^{(0)} \left(\sqrt{M_1^2+\frac{9}{4}}-\frac{1}{2}\right)\right)d M_1^2 \\ \nonumber
 &+& \frac{1}{32 \pi^2}\int^{M_2^2}_{m^2}\left(M_2^2+2\right) \left(\psi^{(0)} \left(\sqrt{M_2^2+\frac{9}{4}}+\frac{3}{2}\right)+\psi^{(0)} \left(\sqrt{M_2^2+\frac{9}{4}}-\frac{1}{2}\right)\right)d M_2^2 \\ \nonumber
 &+& \frac{3}{16 \pi^2}\int^{M_v^2}_{0} \left(M_v^2-2\right) \psi^{(0)} \left(\sqrt{M_v^2+\frac{1}{4}}+\frac{1}{2}\right) d M_v^2 \\ \nonumber
 &-&\frac{1}{32 \pi^2} \int^{M_v^2}_{0}(M^2_v+2) \left(\psi^{(0)} \left(\sqrt{M_v^2+\frac{9}{4}}+\frac{3}{2}\right)+\psi^{(0)} \left(\sqrt{M_v^2+\frac{9}{4}}-\frac{1}{2}\right)\right)d M_v^2 \\ \nonumber
 &-& \frac{1}{64 \pi^2} \phi_{cl}^2 \left(3 \l \left(m^2+2\right) \left(\psi^{(0)} \left(\sqrt{m^2+\frac{9}{4}}+\frac{3}{2}\right)+\psi^{(0)} \left(\sqrt{m^2+\frac{9}{4}}-\frac{1}{2}\right)\right)\right. \\ \nonumber
 &+& \left(m^2+2\right) \left(4 e^2+\l\right)\left(\psi^{(0)} \left(\sqrt{m^2+\frac{9}{4}}+\frac{3}{2}\right)+\psi^{(0)} \left(\sqrt{m^2+\frac{9}{4}}-\frac{1}{2}\right)\right)\\ \nonumber
 &-& \left. 48 e^2  \psi^{(0)} \left(1\right)-8 e^2 \left(\psi ^{(0)}(1)+\psi ^{(0)}\left(3\right)\right)\right)\\ \nonumber
 &-&\frac{1}{768 \pi^2} \phi_{cl}^4\left(27 \l^2  \left(\psi^{(0)} \left(\sqrt{m^2+\frac{9}{4}}+\frac{3}{2}\right)+\psi^{(0)} \left(\sqrt{m^2+\frac{9}{4}}-\frac{1}{2}\right)\right)\right.\\ \nonumber
 &+&\frac{27 \l^2 (m^2+2)}{2 \sqrt{m^2+\frac{9}{4}}} \left(\psi ^{(1)}\left(\sqrt{m^2+\frac{9}{4}}+\frac{3}{2}\right)+\psi ^{(1)}\left(\sqrt{m^2+\frac{9}{4}}-\frac{1}{2}\right)\right) \\ \nonumber
 &+& 3 \left(4 e^2+\l\right)^2 \left(\psi \left(\sqrt{m^2+\frac{9}{4}}+\frac{3}{2}\right)+\psi^{(0)} \left(\sqrt{m^2+\frac{9}{4}}-\frac{1}{2}\right)\right)\\ \nonumber
 &+&\frac{3 (m^2+2) \left(4 e^2+\l\right)^2}{2  \sqrt{m^2+\frac{9}{4}}} \left(\psi ^{(1)}\left(\sqrt{m^2+\frac{9}{4}}+\frac{3}{2}\right)+\psi ^{(1)}\left(\sqrt{m^2+\frac{9}{4}}-\frac{1}{2}\right)\right)\\ \nonumber
 &+&\left. 288 e^4  \psi^{(0)} \left(1\right)- 576 e^4 \psi ^{(1)}\left(1\right)-48 e^4  \left(\psi ^{(0)}(1)+\psi ^{(0)}(3)\right)-32 e^4  \left(\psi ^{(1)}(1)+\psi ^{(1)}(3)\right)\right).
\end{eqnarray}

\noindent

{\bf For $N$ scalars}

\begin{eqnarray}\label{sqedads4N}
 V^0_{eff}&=& m \phi_{cl}^2+\frac{\l \phi_{cl}^4}{4}\\ \nonumber
&+&\frac{2(N-1)}{32 \pi^2}\int^{M_{1i}^2}_{m^2}\left(M_{1i}^2+2\right) \left(\psi^{(0)} \left(\sqrt{M_{1i}^2+\frac{9}{4}}+\frac{3}{2}\right)+\psi^{(0)} \left(\sqrt{M_{1i}^2+\frac{9}{4}}-\frac{1}{2}\right)\right)d M_{1i}^2\\ \nonumber
&+&\frac{1}{32 \pi^2}\int^{M_{1N}^2}_{m^2}\left(M_{1N}^2+2\right) \left(\psi^{(0)} \left(\sqrt{M_{1N}^2+\frac{9}{4}}+\frac{3}{2}\right)+\psi^{(0)} \left(\sqrt{M_{1N}^2+\frac{9}{4}}-\frac{1}{2}\right)\right)d M_{1N}^2 \\ \nonumber
 &+& \frac{1}{32 \pi^2}\int^{M_2^2}_{m^2}\left(M_2^2+2\right) \left(\psi^{(0)} \left(\sqrt{M_2^2+\frac{9}{4}}+\frac{3}{2}\right)+\psi^{(0)} \left(\sqrt{M_2^2+\frac{9}{4}}-\frac{1}{2}\right)\right)d M_2^2 \\ \nonumber
 &+& \frac{3}{16 \pi^2}\int^{M_v^2}_{0} \left(M_v^2-2\right)  \psi^{(0)} \left(\sqrt{M_v^2+\frac{1}{4}}+\frac{1}{2}\right) d M_v^2 \\ \nonumber
 &-&\frac{1}{32 \pi^2} \int^{M_v^2}_{0}(M_v^2+2) \left(\psi^{(0)} \left(\sqrt{M_v^2+\frac{9}{4}}+\frac{3}{2}\right)+\psi^{(0)} \left(\sqrt{M_v^2+\frac{9}{4}}-\frac{1}{2}\right)\right)d M_v^2 \\ \nonumber
 &-& \frac{1}{64 \pi^2} \phi_{cl}^2 \left(2(N-1) \l \left(m^2+2\right) \left(\psi^{(0)} \left(\sqrt{m^2+\frac{9}{4}}+\frac{3}{2}\right)+\psi^{(0)} \left(\sqrt{m^2+\frac{9}{4}}-\frac{1}{2}\right)\right)\right. \\ \nonumber
&+& 3 \l \left(m^2+2\right) \left(\psi^{(0)} \left(\sqrt{m^2+\frac{9}{4}}+\frac{3}{2}\right)+\psi^{(0)} \left(\sqrt{m^2+\frac{9}{4}}-\frac{1}{2}\right)\right) \\ \nonumber
 &+&  \left(m^2+2\right) \left(4 e^2+\l\right)\left(\psi^{(0)} \left(\sqrt{m^2+\frac{9}{4}}+\frac{3}{2}\right)+\psi^{(0)} \left(\sqrt{m^2+\frac{9}{4}}-\frac{1}{2}\right)\right)\\ \nonumber
 &-& \left. 48 e^2  \psi^{(0)} \left(1\right)-8 e^2 \left(\psi ^{(0)}(1)+\psi ^{(0)}\left(3\right)\right)\right)\\ \nonumber
 &-&\frac{1}{768 \pi^2} \phi_{cl}^4\left(
 6(N-1) \l^2  \left(\psi^{(0)} \left(\sqrt{m^2+\frac{9}{4}}+\frac{3}{2}\right)+\psi^{(0)} \left(\sqrt{m^2+\frac{9}{4}}-\frac{1}{2}\right)\right)\right.\\ \nonumber
 &+&\frac{6(N-1) \l^2 (m^2+2)}{2 \sqrt{m^2+\frac{9}{4}}} \left(\psi ^{(1)}\left(\sqrt{m^2+\frac{9}{4}}+\frac{3}{2}\right)+\psi ^{(1)}\left(\sqrt{m^2+\frac{9}{4}}-\frac{1}{2}\right)\right) \\ \nonumber
&+& 27 \l^2  \left(\psi^{(0)} \left(\sqrt{m^2+\frac{9}{4}}+\frac{3}{2}\right)+\psi^{(0)} \left(\sqrt{m^2+\frac{9}{4}}-\frac{1}{2}\right)\right)\\ \nonumber
 &+&\frac{27 \l^2 (m^2+2)}{2 \sqrt{m^2+\frac{9}{4}}} \left(\psi ^{(1)}\left(\sqrt{m^2+\frac{9}{4}}+\frac{3}{2}\right)+\psi ^{(1)}\left(\sqrt{m^2+\frac{9}{4}}-\frac{1}{2}\right)\right) \\ \nonumber
 &+& 3  \left(4 e^2+\l\right)^2 \left(\psi^{(0)} \left(\sqrt{m^2+\frac{9}{4}}+\frac{3}{2}\right)+\psi^{(0)} \left(\sqrt{m^2+\frac{9}{4}}-\frac{1}{2}\right)\right)\\ \nonumber
 &+&\frac{3  (m^2+2) \left(4 e^2+\l\right)^2}{2  \sqrt{m^2+\frac{9}{4}}} \left(\psi ^{(1)}\left(\sqrt{m^2+\frac{9}{4}}+\frac{3}{2}\right)+\psi ^{(1)}\left(\sqrt{m^2+\frac{9}{4}}-\frac{1}{2}\right)\right)\\ \nonumber
 &+&\left. 288 e^4  \psi^{(0)} \left(1\right)- 576 e^4 \psi ^{(1)}\left(1\right)-48 e^4  \left(\psi ^{(0)}(1)+\psi ^{(0)}(3)\right)+32 e^4  \left(\psi ^{(1)}(1)+\psi ^{(1)}(3)\right)\right).
\end{eqnarray}


\begin{thebibliography}{99}




\bibitem{Burgess:1984ti}
C.~P.~Burgess and C.~A.~Lutken,
``Propagators and Effective Potentials in Anti-de Sitter Space,''
Phys. Lett. B \textbf{153} (1985), 137-141
doi:10.1016/0370-2693(85)91415-7

\bibitem{Inami:1985wu}
T.~Inami and H.~Ooguri,
``One Loop Effective Potential in Anti-de Sitter Space,''
Prog. Theor. Phys. \textbf{73} (1985), 1051
doi:10.1143/PTP.73.1051


\bibitem{Inami:1985dj}
T.~Inami and H.~Ooguri,
``NAMBU-GOLDSTONE BOSONS IN CURVED SPACE-TIME,''
Phys. Lett. B \textbf{163} (1985), 101-105
doi:10.1016/0370-2693(85)90201-1

\bibitem{Callan:1989em}
C.~G.~Callan, Jr. and F.~Wilczek,
``INFRARED BEHAVIOR AT NEGATIVE CURVATURE,''
Nucl. Phys. B \textbf{340} (1990), 366-386
doi:10.1016/0550-3213(90)90451-I

\bibitem{Camporesi:1990wm}
R.~Camporesi,
``Harmonic analysis and propagators on homogeneous spaces,''
Phys. Rept. \textbf{196} (1990), 1-134
doi:10.1016/0370-1573(90)90120-Q

\bibitem{Camporesi:1991nw}
R.~Camporesi,
``Zeta function regularization of one loop effective potentials in anti-de Sitter space-time,''
Phys. Rev. D \textbf{43} (1991), 3958-3965
doi:10.1103/PhysRevD.43.3958

\bibitem{Camporesi:1993mz}
R.~Camporesi and A.~Higuchi,
``Arbitrary spin effective potentials in anti-de Sitter space-time,''
Phys. Rev. D \textbf{47} (1993), 3339-3344
doi:10.1103/PhysRevD.47.3339

\bibitem{Camporesi:1994ga}
R.~Camporesi and A.~Higuchi,
``Spectral functions and zeta functions in hyperbolic spaces,''
J. Math. Phys. \textbf{35} (1994), 4217-4246
doi:10.1063/1.530850


\bibitem{Bytsenko:1994bc}
A.~A.~Bytsenko, G.~Cognola, L.~Vanzo and S.~Zerbini,
``Quantum fields and extended objects in space-times with constant curvature spatial section,''
Phys. Rept. \textbf{266} (1996), 1-126
doi:10.1016/0370-1573(95)00053-4
[arXiv:hep-th/9505061 [hep-th]].

\bibitem{Gubser:2002zh}
S.~S.~Gubser and I.~Mitra,
``Double trace operators and one loop vacuum energy in AdS / CFT,''
Phys. Rev. D \textbf{67} (2003), 064018
doi:10.1103/PhysRevD.67.064018
[arXiv:hep-th/0210093 [hep-th]].


\bibitem{Diaz:2007an}
D.~E.~Diaz and H.~Dorn,
``Partition functions and double-trace deformations in AdS/CFT,''
JHEP \textbf{05} (2007), 046
doi:10.1088/1126-6708/2007/05/046
[arXiv:hep-th/0702163 [hep-th]].


\bibitem{Aharony:2010ay}
O.~Aharony, D.~Marolf and M.~Rangamani,
``Conformal field theories in anti-de Sitter space,''
JHEP \textbf{02} (2011), 041
doi:10.1007/JHEP02(2011)041
[arXiv:1011.6144 [hep-th]].

\bibitem{Aharony:2012jf}
O.~Aharony, M.~Berkooz, D.~Tong and S.~Yankielowicz,
``Confinement in Anti-de Sitter Space,''
JHEP \textbf{02} (2013), 076
doi:10.1007/JHEP02(2013)076
[arXiv:1210.5195 [hep-th]].

\bibitem{Kruczenski:2022lot}
M.~Kruczenski, J.~Penedones and B.~C.~van Rees,
``Snowmass White Paper: S-matrix Bootstrap,''
[arXiv:2203.02421 [hep-th]].

\bibitem{Carmi:2018qzm}
D.~Carmi, L.~Di Pietro and S.~Komatsu,
``A Study of Quantum Field Theories in AdS at Finite Coupling,''
JHEP \textbf{01} (2019), 200
doi:10.1007/JHEP01(2019)200
[arXiv:1810.04185 [hep-th]].


\bibitem{Ankur:2023lum}
Ankur, D.~Carmi and L.~Di Pietro,
``Scalar QED in AdS,''
JHEP \textbf{10} (2023), 089
doi:10.1007/JHEP10(2023)089
[arXiv:2306.05551 [hep-th]].

\bibitem{Giombi:2020rmc}
S.~Giombi and H.~Khanchandani,
``CFT in AdS and boundary RG flows,''
JHEP \textbf{11} (2020), 118
doi:10.1007/JHEP11(2020)118
[arXiv:2007.04955 [hep-th]].

\bibitem{Giombi:2021cnr}
S.~Giombi, E.~Helfenberger and H.~Khanchandani,
``Fermions in AdS and Gross-Neveu BCFT,''
JHEP \textbf{07} (2022), 018
doi:10.1007/JHEP07(2022)018
[arXiv:2110.04268 [hep-th]].

\bibitem{Carmi:2019ocp}
D.~Carmi,
``Loops in AdS: From the Spectral Representation to Position Space,''
JHEP \textbf{06} (2020), 049
doi:10.1007/JHEP06(2020)049
[arXiv:1910.14340 [hep-th]].

\bibitem{Carmi:2021dsn}
D.~Carmi,
``Loops in AdS: from the spectral representation to position space. Part II,''
JHEP \textbf{07} (2021), 186
doi:10.1007/JHEP07(2021)186
[arXiv:2104.10500 [hep-th]].


\bibitem{Meineri:2023mps}
M.~Meineri, J.~Penedones and T.~Spirig,
``Renormalization group flows in AdS and the bootstrap program,''
[arXiv:2305.11209 [hep-th]].

\bibitem{Lauria:2023uca}
E.~Lauria, M.~Milam and B.~C.~van Rees,
``Perturbative RG flows in AdS: an \'etude,''
[arXiv:2309.10031 [hep-th]].

\bibitem{Kakkar:2022hub}
A.~Kakkar and S.~Sarkar,
``On partition functions and phases of scalars in AdS,''
JHEP \textbf{07} (2022), 089
doi:10.1007/JHEP07(2022)089
[arXiv:2201.09043 [hep-th]].

\bibitem{Kakkar:2023gzu}
A.~Kakkar and S.~Sarkar,
``Phases of theories with fermions in AdS,''
JHEP \textbf{06} (2023), 009
doi:10.1007/JHEP06(2023)009
[arXiv:2303.02711 [hep-th]].

\bibitem{Gibbons:2006ij}
G.~W.~Gibbons, M.~J.~Perry and C.~N.~Pope,
``Partition functions, the Bekenstein bound and temperature inversion in anti-de Sitter space and its conformal boundary,''
Phys. Rev. D \textbf{74} (2006), 084009
doi:10.1103/PhysRevD.74.084009
[arXiv:hep-th/0606186 [hep-th]].

\bibitem{Giombi:2008vd}
S.~Giombi, A.~Maloney and X.~Yin,
``One-loop Partition Functions of 3D Gravity,''
JHEP \textbf{08} (2008), 007
doi:10.1088/1126-6708/2008/08/007
[arXiv:0804.1773 [hep-th]].

\bibitem{David:2009xg}
J.~R.~David, M.~R.~Gaberdiel and R.~Gopakumar,
``The Heat Kernel on AdS(3) and its Applications,''
JHEP \textbf{04} (2010), 125
doi:10.1007/JHEP04(2010)125
[arXiv:0911.5085 [hep-th]].

\bibitem{Gopakumar:2011qs}
R.~Gopakumar, R.~K.~Gupta and S.~Lal,
``The Heat Kernel on $AdS$,''
JHEP \textbf{11} (2011), 010
doi:10.1007/JHEP11(2011)010
[arXiv:1103.3627 [hep-th]].


\bibitem{Gupta:2012he}
R.~K.~Gupta and S.~Lal,
``Partition Functions for Higher-Spin theories in AdS,''
JHEP \textbf{07} (2012), 071
doi:10.1007/JHEP07(2012)071
[arXiv:1205.1130 [hep-th]].


\bibitem{Denef:2009kn}
F.~Denef, S.~A.~Hartnoll and S.~Sachdev,
``Black hole determinants and quasinormal modes,''
Class. Quant. Grav. \textbf{27} (2010), 125001
doi:10.1088/0264-9381/27/12/125001
[arXiv:0908.2657 [hep-th]].

\bibitem{Martin:2019flv}
V.~L.~Martin and A.~Svesko,
``Normal modes in thermal AdS via the Selberg zeta function,''
SciPost Phys. \textbf{9} (2020), 009
doi:10.21468/SciPostPhys.9.1.009
[arXiv:1910.11913 [hep-th]].

\bibitem{Mueck:1998iz}
W.~Mueck and K.~S.~Viswanathan,
``Conformal field theory correlators from classical field theory on anti-de Sitter space. 2. Vector and spinor fields,''
Phys. Rev. D \textbf{58} (1998), 106006
doi:10.1103/PhysRevD.58.106006
[arXiv:hep-th/9805145 [hep-th]].



\bibitem{Peskin:1995ev}
M.~E.~Peskin and D.~V.~Schroeder,
``An Introduction to quantum field theory,''
ISBN:9780201503975, Addison-Wesley


\bibitem{Coleman:1973jx}
S.~R.~Coleman and E.~J.~Weinberg,
``Radiative Corrections as the Origin of Spontaneous Symmetry Breaking,''
Phys. Rev. D \textbf{7} (1973), 1888-1910
doi:10.1103/PhysRevD.7.1888

\bibitem{Weinberg:1973am}
E.~J.~Weinberg,
``Radiative corrections as the origin of spontaneous symmetry breaking,
[arXiv:hep-th/0507214 [hep-th]]


\bibitem{Tan:1996kz}
P.~N.~Tan, B.~Tekin and Y.~Hosotani,
``Spontaneous symmetry breaking at two loop in 3-D massless scalar electrodynamics,''
Phys. Lett. B \textbf{388} (1996), 611-620
doi:10.1016/S0370-2693(96)01191-4
[arXiv:hep-th/9607233 [hep-th]].

\bibitem{Tan:1997ew}
P.~N.~Tan, B.~Tekin and Y.~Hosotani,
``Maxwell Chern-Simons scalar electrodynamics at two loop,''
Nucl. Phys. B \textbf{502} (1997), 483-515
doi:10.1016/S0550-3213(97)00495-1
[arXiv:hep-th/9703121 [hep-th]].


\bibitem{Shore:1979as}
G.~M.~Shore,
``Radiatively Induced Spontaneous Symmetry Breaking and Phase Transitions in Curved Space-Time,''
Annals Phys. \textbf{128} (1980), 376
doi:10.1016/0003-4916(80)90326-7

\bibitem{Allen:1983dg}
B.~Allen,
``Phase Transitions in de Sitter Space,''
Nucl. Phys. B \textbf{226} (1983), 228-252
doi:10.1016/0550-3213(83)90470-4

\bibitem{Nielsen:1975fs}
N.~K.~Nielsen,
``On the Gauge Dependence of Spontaneous Symmetry Breaking in Gauge Theories,''
Nucl. Phys. B \textbf{101} (1975), 173-188
doi:10.1016/0550-3213(75)90301-6

\bibitem{Fukuda:1975di}
R.~Fukuda and T.~Kugo,
``Gauge Invariance in the Effective Action and Potential,''
Phys. Rev. D \textbf{13} (1976), 3469
doi:10.1103/PhysRevD.13.3469

\bibitem{Patel:2011th}
H.~H.~Patel and M.~J.~Ramsey-Musolf,
``Baryon Washout, Electroweak Phase Transition, and Perturbation Theory,''
JHEP \textbf{07} (2011), 029
doi:10.1007/JHEP07(2011)029
[arXiv:1101.4665 [hep-ph]].

\bibitem{Andreassen:2014eha}
A.~Andreassen, W.~Frost and M.~D.~Schwartz,
``Consistent Use of Effective Potentials,''
Phys. Rev. D \textbf{91} (2015) no.1, 016009
doi:10.1103/PhysRevD.91.016009
[arXiv:1408.0287 [hep-ph]]


\bibitem{Bertan:2018khc}
I.~Bertan and I.~Sachs,
``Loops in Anti\textendash{}de Sitter Space,''
Phys. Rev. Lett. \textbf{121} (2018) no.10, 101601
doi:10.1103/PhysRevLett.121.101601
[arXiv:1804.01880 [hep-th]].

\bibitem{Bertan:2018afl}
I.~Bertan, I.~Sachs and E.~D.~Skvortsov,
``Quantum $\phi^4$ Theory in AdS${}_4$ and its CFT Dual,''
JHEP \textbf{02} (2019), 099
doi:10.1007/JHEP02(2019)099
[arXiv:1810.00907 [hep-th]].


\bibitem{Alday:2020eua}
L.~F.~Alday, M.~Kologlu and A.~Zhiboedov,
``Holographic correlators at finite temperature,''
JHEP \textbf{06} (2021), 082
doi:10.1007/JHEP06(2021)082
[arXiv:2009.10062 [hep-th]].




\end{thebibliography}
\end{document}